\documentstyle[12pt,epsfig]{article}
\begin{document}
\renewcommand{\theequation}{\arabic{section}.\arabic{equation}}
\makeatletter\@addtoreset{equation}{section}\makeatother
\setcounter{secnumdepth}{2}
\begin{titlepage}
\rightline{TAUP 2006-92}
\rightline{November, 1992} \vskip 1in
 
\begin{center}
{\LARGE Pair  production  in  a  strong
electric field: \\
an initial value problem in quantum field
\\theory}\\
\ \\\ \\
{\large Y. Kluger,\footnote{\normalsize Current address:
Theoretical Division T-8, Los Alamos National Laboratory,
Los Alamos, NM 87545, USA} J. M. Eisenberg, and B. Svetitsky\\
\ \\
School of Physics and Astronomy\\
Raymond and Beverly Sackler Faculty\\
of Exact Sciences\\
Tel Aviv University\\
69978 Tel Aviv, Israel\\}
\ \\\ \\
\end{center}
\newpage\noindent
{\em ABSTRACT}. We review recent achievements in the solution of the
initial-value problem for quantum back-reaction in scalar and  spinor  QED.
The problem is  formulated  and  solved  in  the  semiclassical  mean-field
approximation for a homogeneous, time-dependent electric field.  Our
primary motivation in examining back-reaction has to do with  applications
to theoretical models of production of the quark-gluon plasma, though we
here address practicable  solutions  for  back-reaction  in  general.  We
review the application of the method of  adiabatic  regularization  to  the
Klein-Gordon and Dirac fields in order to  renormalize
the expectation value of the current and derive a  finite  coupled  set  of
ordinary differential equations for the time evolution of the  system.
Three  time  scales  are involved in  the  problem  and  therefore   caution
is  needed  to  achieve numerical stability for this system.  Several
physical  features,  like  plasma
oscillations and plateaus in the current, appear in the solution.
From the plateau of the electric current one
can estimate the number of pairs before the onset  of  plasma  oscillations,
while the plasma oscillations themselves yield the number of particles from
the plasma frequency.
 
We  compare the  field-theory  solution  to  a  simple  model  based  on  a
relativistic Boltzmann-Vlasov equation, with a particle  production  source
term  inferred  from  the  Schwinger   particle   creation   rate   and   a
Pauli-blocking (or Bose-enhancement) factor.  This  model  reproduces  very
well the time behavior of the electric  field  and  the  creation  rate  of
charged pairs of the  semiclassical  calculation.   It therefore
provides a simple intuitive understanding of  the nature  of  the  solution
since nearly all the physical  features can be expressed in terms of the
classical distribution function.
\end{titlepage}
 
\section{\bf Introduction}
 
 
This review deals with the physical  situation  in  which  initially  there
exists a classical electric field out of which  pairs of charged particles
tunnel.  The particles
are accelerated by the field, producing a current and,  in
the sequel, a field that acts counter to the original one.  Eventually,
plasma oscillations are set up.  This  so-called
back-reaction problem has been of considerable interest in recent years  in
particle physics, serving for example as a model of the production  of
the  quark-gluon  plasma.  One  supposes   that when  two
nuclei collide at ultra-relativistic energies, they  induce  color
charges on each other.  Passing through each  other, they leave in  their
wake a color electric field from which quarks and gluons emerge
more or less according  to  the  above  scenario.   Of  course,  the  color
electric field is non-Abelian, whereas we take it  to  be  Abelian  in  the
version of the back-reaction problem we defined  above.   Furthermore,  the
field does not fill all space homogeneously, a condition that thus far one
has had to impose in order to carry out  practical  calculations,
as we shall discuss below.
 
Back reaction has been studied intensively in recent years in  the  domain
of  inflationary
cosmology, where instead of a time-varying electric field one has the
time-dependent  gravitational  metric.   Again  pairs  emerge via tunneling
and then act back on the field, in this case through their masses.
 
Our own interest in the back-reaction problem was sparked by its application
to the quark-gluon plasma problem, and  it  is
therefore almost exclusively in these  terms  that  we  shall  present  our
discussion  here.   Much   of   the   evolution   of   regularization   and
renormalization methods for the back-reaction problem [1--17] has taken place,
however, in  the
context of inflationary cosmology. In this  area  there  exists  a  sizable
literature, summarized---except,
of course, for
the more recent developments---in the books of  Birrell  and  Davies
\cite{BD}
and of Fulling \cite{Fulling}.
Earlier surveys of the status of this  problem  can
be found in the reviews of Zel'dovich \cite{Zel72} and of DeWitt
\cite{DeWitt}.  We shall
not describe these developments here, but rather pick up our discussion  with
the  treatment  of   adiabatic   regularization   given   by   Cooper   and
Mottola \cite{CM}, which  has  proved  an  adequate  starting  point  for  the
practicable calculational schemes upon which we focus.
 
In concentrating on methods that lead to workable calculations
for back-reaction, we shall not address the vast  number  of  papers
that deal primarily or exclusively with the  Schwinger  mechanism [22--27]
and  its  many  applications such as the flux-tube
model\footnote{See also \cite{Kogut}.}
[29--32];  nor shall we touch on the
generalizations of this
mechanism to handle pair production in given, external, time-dependent
electric
fields, reviewed in  Refs.~\cite{Marinov77,Grib}; nor do we deal with the
various efforts to consider pair production in restricted volumes
\cite{finvol}.  We
shall restrict ourselves to methods based on adiabatic regularization for the
renormalization of the electric current  or  on  methods  that  have
been shown \cite{Parker74,Birrell78,BD,Fulling}
to  be  equivalent  to  it.\footnote{See  also  the  early
calculation in one spatial dimension by Ambj\o rn and Wolfram
\cite{Ambjorn83}.}
 
Interest in back-reaction in the context of the quark-gluon
plasma\footnote{For
general   introductions   to   this   subject,   see,    e.g.,
\cite{Muller,Hwa}}
(QGP) arose through work of Bia\l as and Czy\.z and coworkers
\cite{BC86,BC88}, of Kajantie and Matsui \cite{Kajantie85}, and
of Gatoff, Kerman, and Matsui \cite{Gatoff87}.
In these series of studies, the authors  modeled
the production and early dynamics of the QGP  by  considering  a  transport
equation for quarks or gluons, on the right-hand side of which  appears  a
source term based on the  Schwinger  formula  for  pair  production  in  an
electric field fixed in time.  The electric field is, in turn, governed  by
Maxwell's equations, with the current of the produced pairs figuring in the
time evolution of the electric field.   This
coupled system is solved,  and the subsequent development exhibits plasma
oscillations  and  other
distinctive behavior (as we shall see in our discussion here).
 
The genesis of
the classical transport equation, while intuitively very appealing, is  far
from self-evident:  The whole point in back-reaction is that  the  electric
field must change in response to the current of produced pairs, so that the
assumption of a fixed field, inherent in the Schwinger formula, is out of
place.  Furthermore,  when  one  imagines  deriving  a  transport  equation
directly from the field equations through the well-known procedure  of  the
Wigner transformation \cite{Dutch}, it
 is clear that the equation cannot lead directly to a source term.\footnote{See
also Refs.~[44--49]
for discussion
of this point, which lies beyond the  scope  of  the  present  review.}
Since the Wigner treatment is completely general, it  follows  that  in  any
quantum mechanical treatment there is no  possible  separation  between  the
role of the electric field in accelerating charged particles and its role in
producing these particles out of the vacuum through tunneling.   Thus  it is
quite mysterious how a classical transport  equation  can  arise   in   this
formalism.
 
In order to examine back-reaction in detail, we return to the original
initial-value problem in quantum field theory.  We write  a  coupled
set of field equations for charged particles in interaction with  an
electric field.
For purposes of comparing with the  models [39--42]
for the QGP, it is adequate to  take  this  electric  field  as  classical.
(Purists may note that this is the leading order in a $1/N$ expansion, where
$N$ is the number of flavors of the matter field.)
Clearly one will wish eventually to consider a quantized
electromagnetic  field (i.e., higher orders in $1/N$)
so as to incorporate the essential physics of radiation for more  realistic
applications.\footnote{There is also the  still  more difficult issue of the
non-Abelian field involved in the QGP; this is also of higher order in
$1/N$.}  The specification of initial
conditions for the matter field reduces the field equations to $c$-number
differential equations.  This development was  carried  out
initially for bosons \cite{CM,PRL}, and is presented here in Section 2.
 
The numerical solution of these coupled equations is confronted by two
obstacles:  (i) The field equations must be renormalized in order to render
them finite, and (ii) it emerges that the calculated momentum distribution
of  the  produced  particles  is  highly
oscillatory so that great care is  required  to  produce
reliable numerical results.  The first  of  these  problems  is  solved  by
taking  as  a  starting  point  the  extensive  work   noted   earlier   in
renormalizing    the    pair-production    problem     for     inflationary
cosmology [1--19], though, as we shall see,  some  modifications  to  this
procedure are required in practice.  Unfortunately, the available techniques
limit us to the case of a spatially homogeneous electric field.
Overcoming the second obstacle merely  requires sufficient
computer power to achieve long-range numerical stability in spite of
short-range oscillations.   These
points are developed in some detail in Section 2.  We there review results
\cite{PRL} in $1+1$ dimensions and present results in $3+1$ dimensions for
the first time.
 
Somewhat to our surprise, the classical transport equation
used [39--42]
for the  QGP  permits very nearly a
quantitative simulation of the field-theory result,  thus  considerably
bolstering one's  belief  in  the  legitimacy  of  various  detailed  steps
involved at the technical  level  in  both  procedures.  The
agreement between the two  procedures  is  greatly  improved  by  including
in the transport equation a term that provides for Bose-Einstein enhancement
in the production process;
this  term  is  found  to  have  a
particular form  which  is  motivated  by  careful  consideration  of  pair
production   in   an   external classical electric field.
 
The treatment of fermion pair production \cite{PRD}, presented in Section 3,
involves some technical issues concerning the  use  of  the
adiabatic regularization procedure.
The physics of Pauli blocking makes  it  rather  less  evident  that  a
classical transport equation can succeed here, so it is, perhaps, even more
surprising than for bosons  that a classical transport  equation
gives an adequate description of  the results,  even  at  a
quantitative level. In this case, good agreement {\it requires}
the introduction of an explicit Pauli-blocking factor in the transport
equation.
 
The application to the QGP
requires going beyond the case of a uniform electric field.
Luckily, one can take advantage of the approximate invariance under
longitudinal boosts which is often assumed in modeling particle production in
the central rapidity region \cite{CooperFrye,Bj83}.
The transformation of the field equations to comoving coordinates
allows renormalization just as in the case of the homogeneous electric field,
and numerical solution follows \cite{biv}.
We do not review these developments here.
 
The work covered here treats a highly-idealized  situation
of a classical electric field, homogeneous in space and producing
particles that possess no mutual interactions.
This can at best serve as
a starting point for more serious studies that  contain  the
complications that are essential for physical applications.   However, we
feel that the
beginning phase is now more or less  complete,  and
therefore is deserving of review here. The incorporation of particle
scattering and of bremsstrahlung is straightforward, and such work is in
progress.
Lifting the restriction of spatial homogeneity (or of boost invariance), on the
other hand, appears very difficult.
 
\section{Boson pair production}
 
\subsection {Introduction}
 
Beginning with Sauter's pioneering paper \cite{Sauter} in 1931, the
problem of spontaneous pair production in the presence of an external
electric field has been investigated by many authors [23--27, 33, 34].
The most commonly used formula for the spontaneous pair creation rate
per unit volume, derived by Schwinger \cite{Schwinger51}, is based on
a field-theory calculation for a constant and homogeneous electric
field $E$, to all orders in $E,$
\begin{equation}
w = (2s+1)\frac{m^4}{(2\pi)^3}\left(\frac{E}{E_0}\right )^2
\sum_{n=1}^{\infty}\frac{\beta_n}{n^2}\exp (-n{\pi}E_0/E)\ ,
\label{PRL_1}
\end{equation}
where $s$ is the spin of the particles produced,
${\beta_n}=(-1)^{n-1}$ for bosons and ${\beta_n}=1$ for fermions, and
$E_0=m^2/e$.
This formula has been used extensively in modeling particle
production in the central rapidity region in high-energy
nucleus--nucleus collisions [39--42, 55, 56].
 
In this expression the electric field is held constant by an external agent.
Thus, there is no effect of the produced particles  on the original electric
field---no back-reaction. Moreover,  the  mutual  interactions  of  the
particles are not taken into account.  As  long  as  the  intensity  of  the
initial electric field is small, $E{\ll}E_0$, Schwinger's formula is  valid.
However,  when it is applied for a strong electric field, $E>E_0$, it  gives
$w\propto (E/E_0)^2$, violating unitarity for sufficiently strong
fields \cite{Marinov77}.  In such
circumstances it  is
necessary to include both
the interaction between the particles created  and  the
screening of the external field.
 
As discussed in Section 1, we assume that the electric field is classical,
Abelian, and homogeneous. We
impose the restriction to a spatially uniform electric field  not  just  for
simplicity, but also to dispose of infinities that would  otherwise  appear
in the expectation value of the current operator in the Maxwell equation.
These
infinities are removed by adiabatic regularization and renormalization.
It is important to note that in an  initial  value  problem  the
divergences in the  expectation  values  of  the  electric  current  or  the
energy-momentum tensor are time dependent. Nonetheless, the charge and  mass
renormalizations are time-independent and the infinities which appear in the
currents  are  products  of  time-independent   infinite   quantities   with
time-dependent finite  quantities.
 
A scheme for solving the quantum back-reaction problem in scalar QED was put
forth by Cooper and Mottola \cite{CM}.
Some analytical progress based on this scheme has been made
\cite{Rogers90}, but is limited to very short times or very weak
electric fields.
In order to see substantial pair production and the subsequent onset
of plasma oscillations we need to begin with a strong initial
electric field, $E\sim E_0$, and to follow the evolution to large
times, $t{\gg}m/eE$.
It is useful to investigate the problem in 1+1 dimensions as a first
step in testing numerical procedures since there are no transverse momenta
to take into account, and since renormalization is
relatively trivial.  We then proceed to $3+1$ dimensions.
 
Adiabatic regularization will only render the theory finite if
we impose a special initial configuration of the  charged matter
field---the adiabatic vacuum.
This does {\em not} mean that there are no particles in the initial state.
As long as the matter field and the electric field interact with each
other, the expectation value of the number operator is {\em not}
equal to the asymptotic particle number density, and even this
adiabatic vacuum may contain a nonzero density.
{\em Nonadiabatic} initial conditions would introduce
infinities into the  time  derivatives  of  the  current  at
$t=0$. Ultimately, the physical behavior we find in our calculations
offers us a way to determine an effective particle density during the
evolution of the system\footnote{One may also add explicitly a finite
density of particles to the initial adiabatic vacuum without
disturbing the finiteness of the theory.}.
 
\subsection{Pair production in semiclassical scalar QED}
 
\subsubsection{Equations of motion and second quantization}
 
The scalar QED coupled equations of motion in the  mean-field  approximation
(and without a $\phi^4$ self-interaction term as  included  in  \cite{CM})
are given by the Klein-Gordon equation
\begin{equation}
[(\partial ^\mu +ieA^\mu )(\partial _\mu +ieA_\mu )+ m^2]\Phi (x)=0
\label{PRL_2}
\end{equation}
and the semiclassical Maxwell equations
\begin {equation}
\partial_\mu F^{\mu \nu}=\langle i\vert j^\nu \vert i \rangle  ,
\label{PRL_3}
\end{equation}
where $\vert i\rangle$ is the initial configuration of the charged matter
field,
and the current $j^\nu$ of the charged scalar field
 is symmetrized in order to be odd under charge conjugation,
\begin{eqnarray*}
j^\nu=\frac{ie}{2}[\Phi^\dagger (D^{\nu} \Phi) -(D^{\nu} \Phi )^{\dagger}\Phi
-\Phi (D^{\nu} \Phi )^\dagger +(D^{\nu} \Phi) \Phi ^{\dagger} ]   .
\label{bose_current}
\end{eqnarray*}
Here $D^{\nu} = \partial^{\nu} + ieA^{\nu}$ is the covariant derivative.
Note that the   expectation
value $\langle i\vert j^\nu \vert  i \rangle$ does  not  vanish  because
$\vert i\rangle$
cannot be an eigenstate of the charge-conjugation operator as long as  a
time-varying electric field is present. As we  shall  see  this  expectation
value is divergent,  and  available  practical  schemes  which  remove  this
divergence are, at the moment, limited to the case of  spatially homogeneous
electric fields \cite{BD,Fulling}.
This homogeneity reduces the Maxwell  equations
to a single equation
\begin{equation}
\frac{\partial ^{2} {\bf A}}{\partial t^2}= \langle i \vert {\bf j}
 \vert i \rangle          .
\label{PRL_4}
\end{equation}
The scalar charge density $j^0$ vanishes everywhere because of Gauss' Law,
and the magnetic  field
is a constant which we choose to be zero. We have selected the gauge $A_0=0$,
and we choose {\bf E} to lie along the $z$ axis.
Thus $A_z$ is the only nonvanishing component, $A^\nu = (0,0,0,A(t))$.
 
Spatial homogeneity enables us to express the boson field operator $\Phi $
as a Fourier integral in which the coefficients are
the Heisenberg operators $a_{\bf k}(t)$ and ${b_{-{\bf k}}}^{\dagger}(t)$,
\begin{equation}
\Phi (x,t)= \int \frac {[d{\bf {k}}]}{\sqrt{2{\omega _{\bf k}}^0}}
\,e^{i{\bf k}\cdot \bf x}\,[a_{\bf k} (t) +{b_{-{\bf k}}}^{\dagger}(t)]\ ,
\label{bose_field}
\end{equation}
where $\omega _{\bf k}^{0} =  \sqrt{{\bf k}^2+  m^2}$  and  $[d{\bf k}]=
{d^d{\bf k}}/{(2\pi)^{d}},$  with  $d$  the  spatial  dimension.  Using   the
Klein-Gordon equation (\ref{PRL_2}), the operator $O_{\bf k} (t)=a_{\bf k}
(t)+ {b_{-{\bf k}}}^{\dagger}(t)$ satisfies
\begin{equation}
\frac{d ^{2} O_{\bf k}}{d t^2}
+\omega _{\bf k}^2 (t)O_{\bf k}=0\ ,
\label{bose_operator}
\end{equation}
where
\begin{equation}
\omega _{{\bf k}} (t)^2 = [{\bf k}-e{\bf{A}}(t)]^2 +m^2\ .
\label{bose_omega}
\end{equation}
Since $\omega _{\bf k}(t)$ is a $c$-number,
the   annihilation   and   creation   operators
$a_{\bf k} (t)$ and  ${b_{-{\bf k}}}^{\dagger}(t)$  can  be  expressed  in
terms of $a_{\bf k} (0)$ and ${b_{-{\bf k}}}^{\dagger}(0)$ by means  of  a
Bogolyubov transformation,
\begin{eqnarray}
a_{\bf k} (t)&=&u_{\bf k}(t)a_{\bf k} (0)+v_{\bf k}(t)
{b_{-{\bf k}}}^{\dagger}(0) \nonumber\\
{b_{-{\bf k}}}^{\dagger}(t)&=&{v_{\bf k}}^\ast (t)a_{\bf k} (0)
+{u_{\bf k}}^{\ast}(t){b_{-{\bf k}}}
^{\dagger}(0).
\label{bose_bogolyubov}
\end{eqnarray}
 
It is convenient to rewrite $\Phi$ in terms of mode amplitudes,
\begin{eqnarray*}
f_{\bf k}(t)=\frac {u_{\bf k}(t)+{v_{\bf k}}^{\ast}(t)}
{\sqrt {2 {\omega _{\bf k}}^0}}\ ,
\end{eqnarray*}
namely,
\begin{equation}
\Phi (x,t) = \int  [d{\bf k}]\,[f_{\bf k}(t)a_{\bf k}(0)+f_{\bf k} ^{\ast}
(t)
b_{-{\bf k}}^{\dagger}(0)]\,e^{i\bf{k} \cdot \bf{x}}\ .
\label{PRL_5}\end{equation}
Thus $f_{\bf k}$ satisfies Eq.~(\ref{bose_operator}),
\begin{equation}
\frac{d^{2} f_{\bf k}(t)}{dt^2}+
\omega _{\bf k}^2 (t)f_{\bf k}(t)=0\ .
\label{PRL_6}
\end{equation}
The Bogolyubov coefficients  are related to the $f_{\bf k} (t)$ by
\begin{equation}
u_{\bf k}(t)=\frac {i\dot{f_{\bf k}} +{\omega _{\bf k}}^0
f_{\bf k}}{\sqrt {2 {\omega _{\bf k}}^0}}
\quad {\rm and} \quad
v_{\bf k}(t)=\frac{i\dot{f_{k}}^{\ast} +\omega _{k}^{0}
f_{k}^{\ast}}{\sqrt{2\omega _{k}^{0}}}\ .
\label{bose_uv}
\end{equation}
The canonical commutation relations
\begin{eqnarray}
[\Phi (t,{\bf x}),\Pi (t,{\bf{y}})]  =  i\delta ^{(d)}(\bf{x}-\bf{y})
\label{bose_commutation}
\end{eqnarray}
imply via (\ref{bose_field}) that
\begin{eqnarray*}
[a_{\bf k}(t), a_{\bf{q}}^{\dagger} (t)]=[b_{\bf k}(t),b_{\bf{q}}^{\dagger}
(t)]
={(2\pi)}^d\delta ^{(d)} (\bf{k}-\bf{q})\ .
\end{eqnarray*}
These relations will be satisfied if
\begin{equation}
{\vert u_{\bf k} (t) \vert } ^2 - {\vert v_{\bf k} (t) \vert }^2 =1\ ,
\label{bose_normuv}
\end{equation}
or, by (\ref{bose_uv}),
\begin{equation}
f_{\bf k}\dot{f}_{\bf k}^{\ast} -f_{\bf k}^{\ast} \dot{f}_{\bf k} = i   .
\label{bose_wronskian}
\end{equation}
Since the Wronskian of Eq.~(\ref{PRL_6}) is time independent, this condition
is satisfied automatically, provided that it is satisfied by the initial
conditions
$f_{\bf k}(0),\dot{f}_{\bf k}(0)$.
 
Eq.~(\ref{bose_wronskian}) implies that $f_{\bf k}(t)$ can be expressed in
 terms of the real and
positive function $\Omega _{\bf k}(t)$,
\begin{equation}
f_{\bf k} (t)=\frac{1}{\sqrt {2 \Omega_{\bf k}(t)}}\,\exp\left[-i\int ^{t}
\Omega_{\bf k}(t^{\prime})dt^{\prime}\right]\ ,
\label{PRL_7}
\end{equation}
if $\Omega_{\bf k}(t)$ satisfies
\begin{equation}
\Omega^{2}_{\bf k}(t) =-\frac{\ddot{\Omega}_{\bf k}}{2\Omega_{\bf k}}
+\frac{3}{4}{\left(\frac{\dot{\Omega}_{\bf k}}{{\Omega_{\bf k}}}\right)}^2
+ \omega^{2}_{\bf k}(t)\ .
\label{PRL_8}
\end{equation}
The representation (\ref{PRL_7}) of $f_{\bf k}$ in a WKB-like form enables
us to obtain  the WKB-like  equation  (\ref{PRL_8})  for  $\Omega_{\bf k}$,
and will help us identify divergences in the
expectation values of the current and the energy-momentum tensor.
 
\subsubsection{The adiabatic expansion and the adiabatic vacuum}
 
The effect of pair production is related \cite{Marinov77} to the appearance
of negative frequencies in the solution for  $f_{\bf k}(t).$
Looked at naively, however,
the parametrization (\ref{PRL_7}) might seem to have only
positive frequencies when the interaction with the electromagnetic
field is switched off.
In fact, this is not the case since
$\Omega _{\bf k}(t)$ for large $t$ does not approach a constant.
To see this, we express $f_{\bf k}(t)$ in a different manner
using the addition formula \cite{Waterman73}
\begin{eqnarray}
f_{\bf k} (t)&=&\frac{1}{\sqrt {2 \Omega_{\bf k}(t)}}\,\exp\left[-i\int ^{t}
\Omega_{\bf k}(t^{\prime})dt^{\prime}\right] \nonumber\\
&=& \frac{1}{\sqrt {2 W_{\bf k}(t)}}\left\{c_{1{\bf k}}\,
\exp\left[-i\int ^{t}
W_{\bf k}(t^{\prime}) dt^{\prime}\right] \right.\nonumber\\
&&\left.\qquad\qquad\mbox{}+
c_{2{\bf k}}\,\exp\left[+i\int  ^{t}W_{\bf k}(t^{\prime})dt^{\prime}\right]
\right\}\ ,
\label{bose_addition}
\end{eqnarray}
where the constants $c_{1{\bf k}}$ and $c_{2{\bf k}}$ fulfill
\begin{equation}
{\vert c_{1{\bf k}}\vert } ^2 - {\vert c_{2{\bf k}} \vert }^2 =1
\label{bose_normc1c2}
\end{equation}
and where $W_{\bf k}(t)$ satisfies the same equation of motion (\ref{PRL_8})
as does $\Omega_{\bf k}(t)$.
According to a theorem in asymptotic expansions, which  is  related  to  the
generalized  WKB  or   Liouville-Green   approximation,   if   $\omega   \in
C^{\infty}$, the solutions of (\ref{PRL_6}) are linear combinations of basis
solutions [the  right-hand  side  of  (\ref{bose_addition})]  which  can  be
approximated as \cite[p.~154]{Fulling}
\begin{eqnarray}
f_{\bf k}^{\pm} (t)
& \equiv & \frac{1}{\sqrt {2 W_{\bf k}(t)}}
\,\exp\left[\mp i\int ^{t} W_{\bf k}(t^{\prime}) dt^{\prime}\right]
\nonumber\\
&=&\frac{1}{\sqrt {2 W^{\rm ad}_{\bf k}(t)}}
\,\exp\left[\mp i\int ^{t} W^{\rm ad}_{\bf k}(t^{\prime}) dt^{\prime}\right]
+O(\omega^{-N}_{\bf k}) \nonumber \\
W^{\rm ad}_{\bf k}(t)^2 & = & \omega_{\bf k} (t)^2
[1+\delta_{2{\bf k}} (t)\omega^{-2}_{\bf k}+
\delta_{4{\bf k}} (t)\omega^{-4}_{\bf k} + ...].
\label{bose_Fulling_7_16}
\end{eqnarray}
Here  $\delta_{n{\bf k}}$  is  a  function  of  $\omega_{\bf k}$  and  its
derivatives    at     $t$     up     through     $\omega^{(n)}_{\bf k}(t)$,
and
$N$ is determined by the order at which the series is truncated.
The coefficients
$\delta_{n{\bf k}}$ are bounded as $\omega_{\bf k} \rightarrow \infty$.
Furthermore,
the error term is uniform in $t$ on finite intervals. In  a  static  region,
when the interaction is switched  off  (as  is  expected  at  large  times),
$\delta_{n{\bf k}}=0$. Therefore, it is obvious that the  general  solution
$f_{\bf k}(t)$  of  (\ref{PRL_6}), where both
$c_{1{\bf k}},c_{2{\bf k}}  \neq  0$,
exhibits negative frequencies as  well  as  positive  ones.
 
The amplitude
$f_{\bf k}(t)$ can also be represented by a combination of the WKB solution
and its complex conjugate with time-dependent coefficients
\begin{eqnarray}
f_{\bf k} (t)=\frac{1}{\sqrt {2 W^{\rm ad}_{\bf k}(t)}}
\left[\alpha _{\bf k} (t) e^{-i\int ^{t} W^{\rm ad}_{\bf k}(t^{\prime})
dt^{\prime}}+
\beta _{\bf k}(t) e^{+i\int ^{t}W^{\rm ad}_{\bf k}(t^{\prime})dt^{\prime}}
\right],
\label{bose_f_alphabeta}
\end{eqnarray}
where $W^{\rm ad}_{\bf k}$ is the generalized WKB function, up to the desired
order.  For large momentum  ${\bf k}$, or  at
times when the electromagnetic interaction is absent, we have
$W_{\bf k}(t)  \approx
W^{\rm ad}_{\bf k}(t)$. In this adiabatic limit  $W^{\rm ad}_{\bf k}(t)$ is
nearly
a constant, and $\alpha_{\bf k}$ and $\beta_{\bf k}$  must be constant  up
to the adiabatic order  $N$  to  which  we  expanded  $W^{\rm ad}_{\bf k}(t)$,
because $\left[2 W^{\rm ad}_{\bf k}(t)\right]^{1/2}\,\exp\left(\mp i\int^{t}
W^{\rm ad}_{\bf k}(t') dt'\right)$ are solutions of (\ref{PRL_6})  to
this    order.     If     we     choose     $\alpha_{\bf k}(t_0)=1$     and
$\beta_{\bf k}(t_0)=0$ for an initial time $t_0$ it follows that
\begin{eqnarray}
\alpha_{\bf k}(t)=1+O(\omega^{-N-1}_{\bf k}) \quad {\rm and} \quad
\beta_{\bf k}(t)=0+O(\omega^{-N-1}_{\bf k})
\label{bose_alphabeta}
\end{eqnarray}
for all times. The exact modes $f_{\bf k}$ which correspond to this  choice
are denoted ``adiabatic positive-frequency  modes of adiabatic
order $N$.''    Once     we     choose     $\alpha_{\bf k}(t_0)=1$     and
$\beta_{\bf k}(t_0)=0$,  then,  for  large  momentum,  $c_{1{\bf k}}$  and
$c_{2{\bf k}}$ are
\begin{eqnarray}
c_{1{\bf k}}=1+O(\omega^{-N-1}_{\bf k}) \quad {\rm and} \quad
c_{2{\bf k}}=0+O(\omega^{-N-1}_{\bf k}).
\label{bose_c1c2}
\end{eqnarray}
Note that at  $t=t_0$  the  exact  modes  are  equal to  their  adiabatic
approximation of order $N$. As is seen  from  the  above  theorem,  the  WKB
approximation for $W_{\bf k}$ is asymptotically correct.
 
We see above that the   exact  adiabatic  positive-frequency  modes  can  be
written  as  a  linear  combination  (\ref{bose_addition}),   and  for   our
particular  choice,  \mbox{$\alpha_{\bf k}(t_0)=1$}  and  
$\beta_{\bf k}(t_0)=0$,
condition (\ref{bose_c1c2}) is satisfied. An asymptotic  expansion of modes
$\Omega_{\bf k}$ may therefore differ from the  expansion  of  $W_{\bf k}$
only in order $O(\omega^{-N-1}_{\bf k})$. Thus it is  legitimate to extract
the large ${\bf k}$ behavior of $\Omega_{\bf k}$ by  adiabatic  expansion,
even  though  $\Omega_{\bf k}$  contains  elements  of  both  positive  and
negative frequencies. We also equate the time derivatives of the exact modes
with the time derivatives of the $N$th order approximate adiabatic modes  at
$t=t_0$. The  annihilation  operators  $a_{\bf k}$  and  $b_{\bf k}$  that
correspond to these modes (which are  matched  with  their  $N$th  adiabatic
approximations at $t=t_0$) annihilate a vacuum state, which is  said  to  be
the ``adiabatic vacuum'' $\vert 0\rangle_A$. If $\omega_{\bf k}$         is
asymptotically static, then $f_{\bf k}^{\rm in}$ and $f_{\bf k}^{\rm out}$
are  in
the  class  of solutions approximated by the positive-frequency modes
$f_{\bf k}^{+}$  for any $N$. In this special case the adiabatic vacuum can
be  identified  with the usual vacua $\vert 0^{\rm in}\rangle$ and $\vert
0^{\rm out}\rangle$,  where  the  exact modes are matched  to  their  adiabatic
approximation  at  $t \to -\infty$ and $t \to +\infty$, respectively.
 
As we shall see, our initial conditions for the  back-reaction  problem  are
not static because of the presence of an electric field at  $t=t_0$.  It  is
important to note that this adiabatic  vacuum  is  not  precisely  a  vacuum
state, that is, an inertial particle  detector  would  register  a  bath  of
quanta when the field is in this adiabatic vacuum. Furthermore, the matching
of the mode functions may be specified at  any  time  $t_0$,  and  therefore
there is no unique $N^{\rm th}$ order adiabatic vacuum.
Thus  there  is  always  an
inherent ambiguity in the definition of the particle concept, and as long as
the rate of particle production is not negligible, the  notion  of  particle
number is not useful.  Nevertheless,
the adiabatic vacuum is the best candidate to
represent the vacuum  for  problems  that  are  not  asymptotically  static
because there is a large probability that it leaves  the  high-energy  modes
empty. This will prove to be crucial, since it  enables  the  isolation  and
regularization of the infinities which appear in the expectation  values  of
the currents. A correct way to describe the dynamical evolution of  the  state
of  a  field  is  to  calculate  the  evolution  of  local field
observables, such as  $\langle  \psi\vert j^{\mu}(x)\vert\psi  \rangle$  or
$\langle\psi\vert T^{\mu \nu}(x)\vert\psi \rangle$, instead of  global
quantities like  the particle number.
 
\subsubsection{The  expectation  value  of   the   current   and   adiabatic
regularization}
 
We now turn to the  calculation  of  the
current expectation value.  As a result of spatial homogeneity we have
\begin{eqnarray}
\langle a^{\dagger}_{\bf k}a_{\bf{q}} \rangle &=&{(2\pi)}^{d}\delta ^{d}
({\bf{k-q}})N_{+}({\bf k})\nonumber\\
\langle b^{\dagger}_{-{\bf k}}b_{-{\bf{q}}} \rangle &=&{(2\pi)}^{d}
\delta ^{d}({\bf{k-q}})N_{-}({\bf k})
\label{CM_2_24}\\
\langle b_{-{\bf k}}a_{\bf{q}} \rangle &=&{(2\pi)}^{d}
\delta ^{d}({\bf{k-q}})F({\bf k})\nonumber
\end{eqnarray}
where $N_{+}$ and $N_{-}$ are the  positive and negative charge  densities  in
the initial state and $F({\bf k})$ is the correlation pair density or
fluctuation term. Gauss' Law dictates
$N_{+}=N_{-}\equiv N$, since
\begin{equation}
\langle j^0 \rangle =e \int [d{\bf k}][N_{+}({\bf k})-N_{-}({\bf k})]=0.
\label{CM_2_25}
\end{equation}
Using   (\ref{PRL_4}),   (\ref{PRL_5}),   (\ref{PRL_7}),   and
(\ref{CM_2_24}), we have
\begin{eqnarray}
\ddot{A} &=&\langle j^3\rangle\nonumber\\
&=&e\int [d{\bf k}]\frac{(k^3 -eA)}{\Omega_{\bf k}}\left
 [1+2N({\bf k}) + 2F({\bf k})\cos \left( 2\int ^{t}\Omega_{{\bf k}}(t^{\prime})
 dt^{\prime}\right)\right].\nonumber\\
&& \label{CM_3_12}
\end{eqnarray}
Here $N({\bf k}) \geq 0$  and $F({\bf k})$ can be chosen to be  real  with
$F({\bf k}) \geq 0$ by adjusting the overall phase of the  mode  amplitudes
$f_{\bf k}(t)$.   The  expression in the square brackets in
(\ref{CM_3_12}) is positive. To prove this, we note
\begin{equation}
0\leq\langle (a^{\dagger}_{\bf k}-b_{-{\bf k}})
(a_{\bf k}-b^{\dagger}_{-{\bf k}})\rangle=
1+2N({\bf k})-2F({\bf k})\ ,
\label{CM_2_31}
\end{equation}
and hence
\begin{equation}
0\le 2F({\bf k}) \le 2N({\bf k})+1\ .
\label{CM_2_32}
\end{equation}
By inspecting  the  brackets  of
(\ref{CM_3_12}), we can identify the first term as the vacuum  contribution,
the second term as roughly the on-shell classical part of the  current,  and
the last term  as  the  quantum  fluctuation  part,  which  oscillates  with
frequencies larger than $2m$.
 
Because the densities of particles  and  correlated  pairs
at $t=0$ are finite, only the vacuum term
\begin{equation}
e\int [d{\bf {k}}](k^3 -eA)\frac{1}{\Omega_{\bf k}}
\label{CM_3_13}
\end{equation}
contains infinities.
In 1+1 dimensions, there is only a normal-ordering infinity in the current,
which is easily subtracted.
In 3+1 dimensions, an infinite charge renormalization is needed as well.
These divergences can be identified and subtracted with adiabatic
regularization.\footnote{It  may  be shown
\cite{Parker74,Birrell78,Bunch80,BD,Fulling}
that all the divergences obtained in adiabatic regularization are the same as
those found via point-splitting  regularization or via the closely related
Pauli-Villars \cite{BD}
and $n$-wave \cite{Zel71,Parker74} schemes.}
 
In order to isolate the divergent part of (\ref{CM_3_12}), we examine the
large-momentum behavior of $\Omega_{\bf k}$.
It is convenient to expand $1/\Omega_{\bf k}$ in powers of
$1/\omega_{\bf k}$.  (As discussed above, this asymptotic expansion is valid
for positive-frequency modes.)
This is done by  successive  iteration of the mode equation (\ref{PRL_8}):
Inserting  the  zeroth-order
solution   $\Omega^{(0)}_{{\bf k}s}=\omega_{\bf k}$  into  the  right-hand
side of (\ref{PRL_8}) one obtains $\Omega^2$ up to second  order;  inserting
this value in the right-hand side the fourth order is obtained, and  so  on.
It is not  difficult  to  see  that  higher-order  adiabatic  approximations
contain terms of higher order in  $1/\omega_{\bf k}$.  We  shall  see  that
adiabatic expansion  up  to  second  order  is  sufficient  to  isolate  the
divergences in the current.  Noting that
\begin{equation}
\ddot{\omega}_{\bf k}=\frac {-e\ddot{A}(k^3-eA)}{\omega_{\bf k}}
+O(1/ \omega_{\bf k})\ ,
\label{CM_3_18}
\end{equation}
we have up to second order
\begin{equation}
\frac{1}{\Omega_{{\bf k}}(t)}=\frac{1}{\omega _{\bf k}}-\frac{e\ddot{A}
(k_{3}-eA)}{4\omega^{5}_{{\bf k}}(t)}
+O(1/ \omega^5_{\bf k})\ .
\label{CM_3_15}
\end{equation}
 
Formally we can write the integrand of (\ref{CM_3_12}) as
\begin{equation}
\frac{1}{\Omega_{\bf k}}\left [1+2N(k)+
2F(k)\cos\left( 2\int ^{t}\Omega_{k}(t^{\prime}) dt^{\prime} \right )\right ]=
\frac{1}{\omega _{\bf k}}-\frac{e\ddot{A}
(k_{3}-eA)}{4\omega^{5}_{{\bf k}}(t)} + r_{\bf k}(t),
\label{CM_4_3}
\end{equation}
where $r$ is a remainder term which falls off faster than $k^{-4}$ for
${\bf k} \rightarrow \infty $.
By substituting (\ref{CM_4_3}) into (\ref{CM_3_12}) and replacing $k^3
 -eA$ by $k^3$ in
the second term of the integral we find
\begin{equation}
\ddot{A}=e\int [d{\bf k}]\frac{1}{\omega _{{\bf k}}(t)}(k^3-eA)-
e^2\delta e^2\ddot{A}+e\int [d{\bf k}](k^3 -eA)r_{\bf k}(t)\ ,
\label{bose_ddA}
\end{equation}
where
\begin{equation}
\delta e^2 \equiv \frac{1}{4}\int [d{\bf k}]
\frac{k^{2}_{3}}{(\omega^0_{\bf k})^5}
=\frac{1}{4d} \int [d{\bf k}] \frac{{\bf k}^{2}}{(\omega^0_{\bf k})^5}\ .
\label{bose_delta_e_square}
\end{equation}
The first term vanishes upon symmetric integration, while
the second term is the same as the divergent part of
the scalar vacuum polarization $\Pi(q^2=0)$, which is logarithmically
divergent in 3+1 dimensions.
 
Note that shifting the  variable  of  integration  in  the  first  term  of
(\ref{bose_ddA}) is not trivial, because the integral is linearly divergent
when $d=1$ and cubically divergent when $d=3$.
A correct, gauge invariant way to perform the integration and discard this
infinity is by cutting off $k^i$ at
$\mp \Lambda +eA^i$, which corresponds to fixed integration boundaries for  the
kinetic momentum ${\bf k}-e{\bf{A}}$. This automatically gives the result
of symmetric integration.
Alternatively, the infinity  can  be
removed by Pauli-Villars regularization, adding  one  auxiliary  massive
field to the theory.
 
After removing the first term, we define the renormalized charge and gauge
field via
\begin{equation}
e^{2}_{R}=e^{2}(1+e^2\delta e^2)^{-1}\equiv
Ze^2  ,\qquad A_R=Z^{-1/2}A\,,
\label{CM_3_22}
\end{equation}
and multiply (\ref{CM_3_12}) by the factor $Ze/e_R$. The renormalized
equation is
\begin{eqnarray}
\ddot{A}_R &=& e_R\int [d{\bf k}]\frac{(k^3 -e_R A_R)}{\Omega_{\bf k}}
\left[1+2N({\bf k})+2F({\bf k})\cos
\left(2\int ^{t}\Omega_{{\bf k}}(t^{\prime})
dt^{\prime}\right)\right] \nonumber \\
&&+ e_R^2\ddot{A}_R\delta e^2 .
\label{CM_4_4}
\end{eqnarray}
The last term formally cancels the divergence of the first.
This can be seen explicitly by using (\ref{CM_4_3}) in (\ref{CM_4_4}), giving
\begin{equation}
\ddot{A_{R}}=e_{R}\int [d{\bf k}](k^3 -e_{R}A_{R})r_{\bf k}(t)
\label{CM_jR}
\end{equation}
in terms of the renormalized quantities. Superficially,  it  seems  that
the second  derivative  of  $A$  appears  only  on  the  left-hand  side  of
(\ref{CM_jR}), but in fact the subsidiary condition (\ref{CM_4_3})  defining
$r_{\bf k}$ is an intrinsic part of  (\ref{CM_jR}).
 
Since  $eA=e_{R}A_{R}$
and  since  $r({\bf k},t)$  depends  on  $e$  and  $A$  only  through   the
combination $eA$, we will henceforth drop the subscript $R$.
 
\subsubsection{Practicalities of numerical solution}
 
Now that we have renormalized the equations we turn to the  practical  aspects
of solving the back-reaction  problem.  We  are  interested  in  solving  an
initial-value problem.
The initial conditions for the Maxwell equation
are
\begin{equation}
\dot{A}(t=0) = -E_0\ ,\quad  A(t=0)=0\   .
\label{CM_4_2}
\end{equation}
The initial state of the boson field
is the adiabatic vacuum,  selected  by  matching
the exact solutions of (\ref{PRL_6})  to  their  adiabatic  approximation,
viz.,
\begin{equation}
\Omega _{{\bf k}}(t=0) = \omega_{{\bf k}}(t=0)\ , \quad
\dot{\Omega} _{{\bf k}}(t=0) = \dot{\omega}_{{\bf k}}(t=0)\ ,
\label{bose_initialmodes}
\end{equation}
and by setting
\begin{equation}
N({\bf k})=F({\bf k})=0\ .
\label{bose_initialNF}
\end{equation}
Nonvacuum  initial   conditions   may   be specified by   adding   nonzero
particle-number and pair-correlation densities to  the  current  expectation
value, without changing the  initial  conditions  (\ref{bose_initialmodes}).
It is important to note that (\ref{CM_4_3}), in which $r_{\bf k}$ must fall
off faster that $k^{-4}$ at large $k$, constrains the initial conditions
placed on $\Omega_{\bf k}$.
The last two terms in  (\ref{CM_4_3}) are connected at the same time by the
Maxwell equation (\ref{CM_jR}).
To check the consistency of the initial conditions (\ref{bose_initialmodes}),
we begin by noting that requiring $\Omega_{\bf k}=\omega_{\bf k}$ turns
(\ref{CM_4_3}) into
\begin{equation}
0=-\frac{e\ddot{A}
(k_{3}-eA)}{4\omega^{5}_{{\bf k}}(t)} + r_{\bf k}(t)\ ,
\label{bose_consistency}
\end{equation}
which is a homogeneous equation in $\ddot A$ and $r_{\bf k}$, as is the
Maxwell equation (\ref{CM_jR}).
The solution to the system (\ref{bose_consistency}) and (\ref{CM_jR})
is, obviously, $\ddot A=r_{\bf k}=0$.
The consistency test comes from differentiating (\ref{CM_4_3})
and (\ref{CM_jR}) with respect to time.
Upon inserting the initial conditions (\ref{CM_4_2}) and
(\ref{bose_initialmodes}), the two equations are homogeneous in $d^3\!A/dt^3$
and $\dot r_{\bf k}$, and possess the consistent solution $d^3\!A/dt^3=
\dot r_{\bf k}=0.$
 
Given such a set of consistent initial conditions, one can evolve the set of
back-reaction equations (\ref{PRL_8}), (\ref{CM_4_3}),   and  (\ref{CM_jR}).
By  stepping   the  equations  forward  in  time,   one   arrives   at   $\{
A,\dot{A},\Omega_{{\bf k}},\dot{\Omega}_{{\bf k}} \}$  at  each  time.  In
principle, $\ddot{A}$ can be calculated without using (\ref{CM_jR}). One can
examine the asymptotic behavior of $\Omega ({\bf k})$ at large momentum and
extract both $r_{\bf k}$ and $\ddot{A}$. For instance, for large ${\bf k}$
one can parametrize the left-hand side of (\ref{CM_4_3}) as
\begin{equation}
\frac{1}{\Omega_{\bf k}}=\frac{1}{\omega _{\bf k}}-\frac{e\ddot{A}
(k_{3}-eA)}{4\omega^{5}_{{\bf k}}(t)}
+\frac{\alpha}{\omega^5_{\bf k}}
+\frac{\beta}{\omega^6_{\bf k}}+ \cdots ,
\label{bose_fit}
\end{equation}
(for $N=F=0$)
and directly extract the value of $\ddot{A}$. The fitted value of $\ddot{A}$
depends, however, on the number of terms used to expand the above series. In
addition, at  some  momentum,  a  fit  to  a  finite  series  in  powers  of
$\omega^{-1}_{\bf k}$ breaks down. Practically one faces another difficulty
in solving  the  semiclassical  system  (\ref{PRL_8}),  (\ref{CM_4_3}),  and
(\ref{CM_jR}):  A numerical solution of  the mode equation using  a  standard
scheme, such as Runge-Kutta, does not necessarily  agree  with  the  correct
adiabatic expansion for large momentum, simply because the Taylor  expansion
in the time parameter does not converge for large ${\bf k}$.
 
Instead, we adopt an iterative scheme making use of (\ref{CM_jR}). We take
$r_{\bf k}$ to be zero at an extremely large momentum as a trial value,  so
that $\ddot{A}$ is extracted from (\ref{CM_4_3}) automatically.  With this
value for $\ddot{A}$ we use (\ref{CM_4_3}) to extract $r_{\bf k}$ for  each
${\bf k}$  up  to  this  very  large  momentum.   Then,   by   substituting
$r_{\bf k}$ in (\ref{CM_jR}) we get a new and slightly different value  for
$\ddot{A}$. This procedure may be iterated until convergence of the sequence
of ($\ddot{A}$, $r_{\bf k}$) is reached. Thereafter the  next  time  step  is
taken.
 
\subsection {Numerical results in (1+1) dimensions}
 
It is useful to investigate the set of  coupled   Klein-Gordon  and  Maxwell
equations  in  1+1  dimensions,  since  the momentum integral in (\ref{CM_jR})
is only one-dimensional and thus the computations are more manageable.
The renormalization procedure is somewhat simpler, since
the  charge  renormalization  is  finite   [see
(\ref{bose_delta_e_square})].  Returning to (\ref{CM_4_4}), we can now solve
for $\ddot A$, obtaining
\begin{equation}
\ddot{A}_R=\langle j^1\rangle =
\frac{e_R}{1-e^2_R\delta e^2}\int \frac{dk}{2\pi}(k-e_R A_R)
\left [\frac{1}{\Omega_k(t)} - \frac{1}{\omega_k (t)} \right ] .
\label{PRL_9}
\end{equation}
The subtraction of $1/\omega_k(t)$ in the integrand has been inserted
by hand.
Just like the apparent divergence in the unsubtracted integral, the integrated
subtraction also vanishes upon symmetric integration.
This term is inserted to ensure convergence, irrespective of how the numerical
integration treats the large $|k|$ regions.
Eq.~(\ref{PRL_9}) is to be coupled to the one-dimensional version of
(\ref{PRL_8}),
\begin{equation}
\Omega^2_k(t) =-\frac{\ddot{\Omega}_k}{2\Omega_k}
+\frac{3}{4}{\left(\frac{\dot{\Omega}_k}{\Omega_k}\right)}^2
+ \omega^{2}_k(t)\ .
\label{PRL_81d}
\end{equation}
The initial conditions are given by
(\ref{CM_4_2})--(\ref{bose_initialNF}).
 
\begin{figure}[htb]
\epsfig{width=6.5cm,file=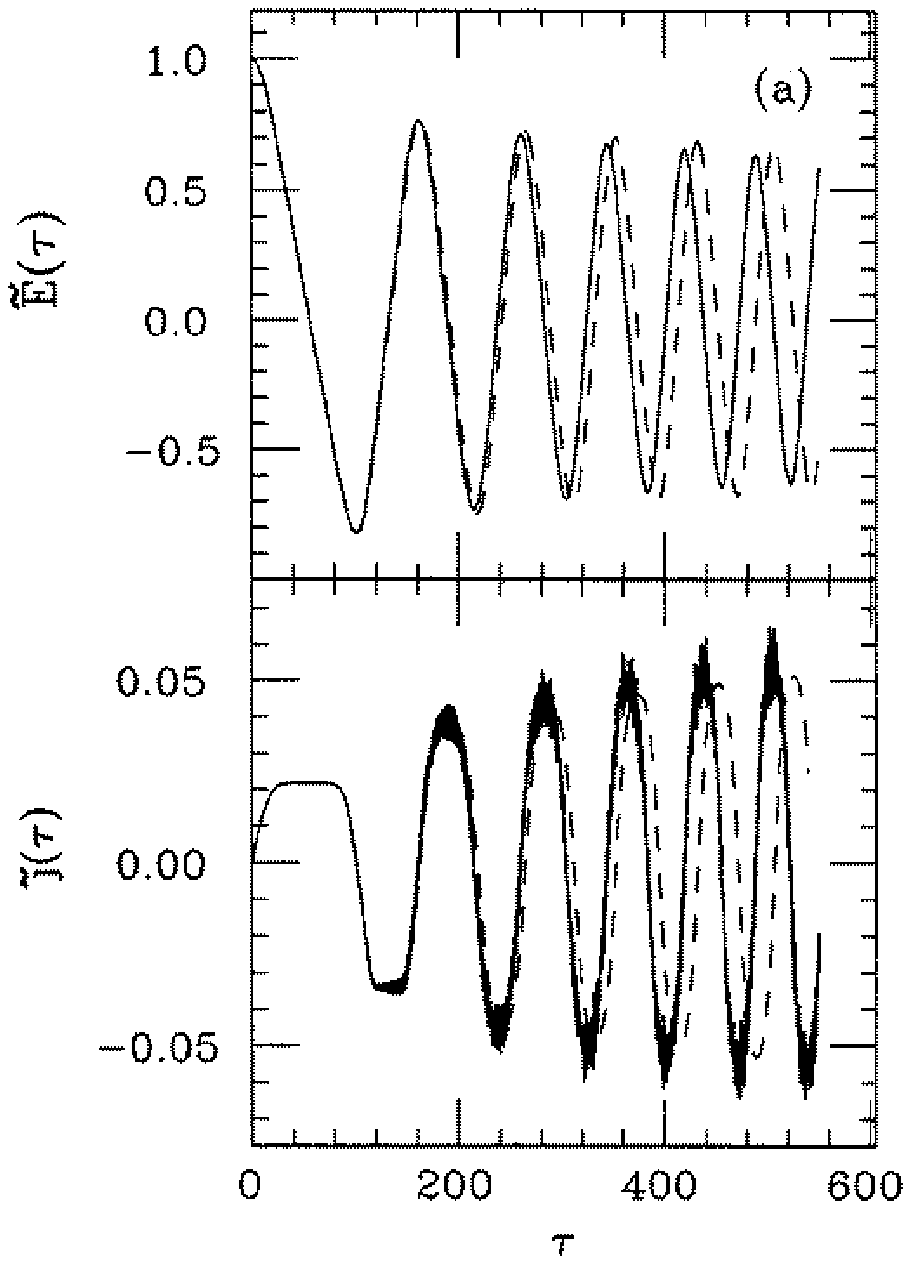}
\epsfig{width=6.25cm,file=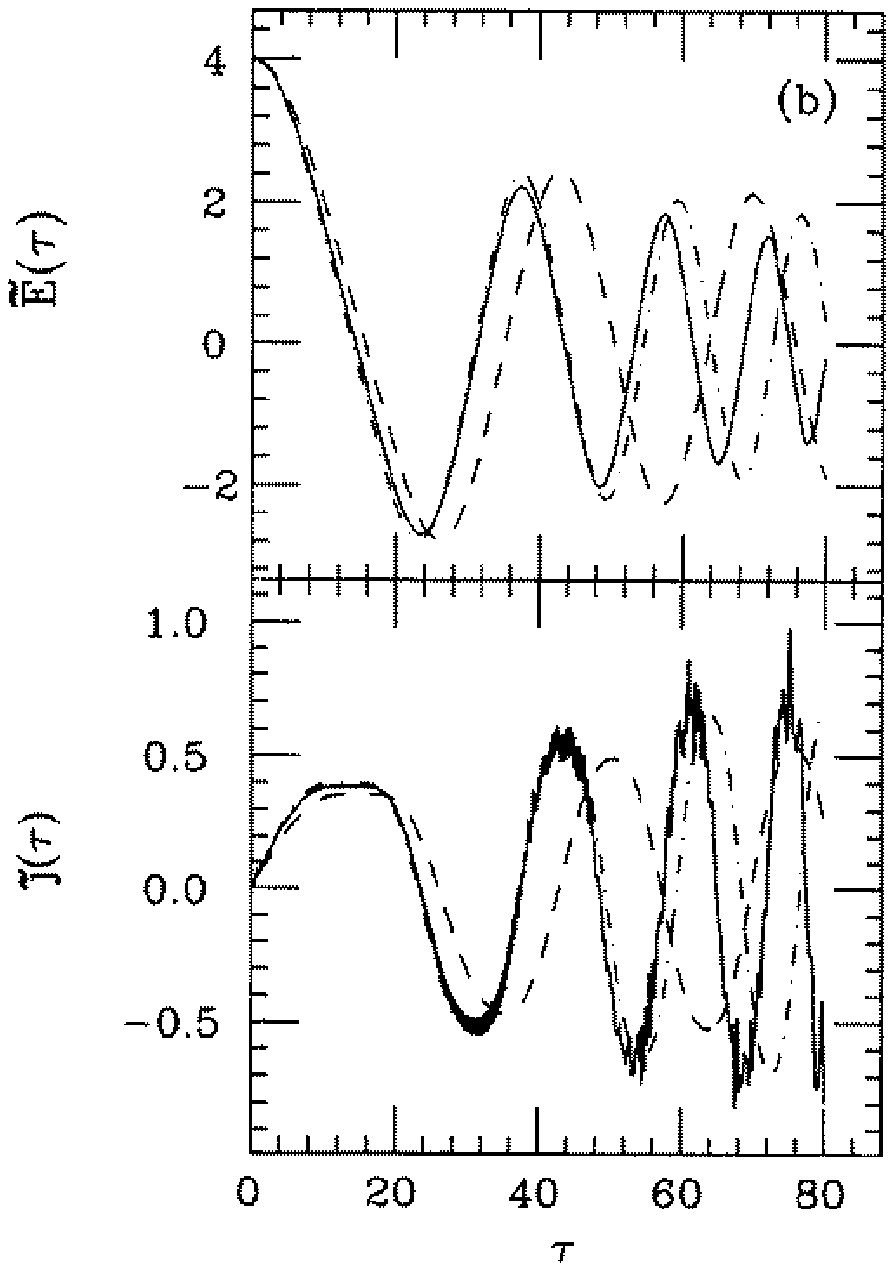}
\caption{(a) Time evolution  of  scaled  electric  field  $\tilde  E$  and
current $\tilde \jmath$, with initial  value  $\tilde  E=1.0$  and  coupling
$e^2/m^2=0.1$. Solid line is semiclassical scalar QED, and  dashed  line  is
Boltzmann-Vlasov model {\em without} the  stimulated  pair
creation correction, Eq. (\ref{PRL_11a}).
(b) Same  as  (a),  but  for  $\tilde  E=4.0$.   The dashed curve is
as in (a), while the
dashed-dotted curve {\em includes} the correction, Eq. (\ref{PRL_17}).
\label{fig2_1}}
\end{figure}

We show in Fig.~\ref{fig2_1} the time evolution of the scaled electric field
\break\mbox{$\tilde E\equiv  eE/m^2$}
and  induced  current  $\tilde  \jmath\equiv  ej/m^3$,  as
functions of $\tau\equiv mt$.
With strong initial  electric  fields  we  find  that  the  induced
current increases rapidly and becomes saturated at a constant value for some
time, after which plasma oscillations are clearly seen.
The fine oscillations in $\tilde\jmath$ are {\it not} numerical artifacts,
since
they persist with small time steps ($d\tau=10^{-4}$) and fine momentum grids
($d\tilde k\equiv dk/m=0.004$).
 
In order to have some insight into these results, we  examine  an  analogous
back-reaction problem of a classical system of particles  and  antiparticles
which interact with a homogeneous electric field. For simplicity, we  assume
that the momentum distributions $N^\pm$ of the particles and the antiparticles
are localized at $\pm p'$,
\begin{eqnarray}
N^\pm(p,x)=2\pi\delta(p\mp p')n\ ,
\label{bose_N(p)}
\end{eqnarray}
where $n$ is the density of particles and $p$ is the kinetic momentum.
The set of coupled equations for this system reads
\begin{eqnarray}
\ddot{A} &=&  j= 2en v=2en\frac{p}{\sqrt{p^2+m^2}} \nonumber \\
\dot{p} &=&  eE=-e\dot{A}\ ,
\label{bose_classical}
\end{eqnarray}
where
\begin{equation}
p = \gamma m v, \quad
\gamma = (1-{v}^2)^{-1/2}.
\label{bose_p_gamma}
\end{equation}
The factor 2 in the current accounts for the antiparticles. This system  has
an integral of motion  from energy conservation,
\begin{equation}
\epsilon=\frac{1}{2}E^2+2n\sqrt{p^2+m^2}
\label{bose_integral_motion}
\end{equation}
and thus
\begin{equation}
\epsilon=\frac{\dot{p}^2}{2e^2}+2n\sqrt{p^2+m^2}\ ,
\label{bose_clasic_energy}
\end{equation}
where $\epsilon$ is a constant. The  last  equation  is  equivalent  to  the
system (\ref{bose_classical}),  and  it  reduces  the  system  to the single
equation
\begin{equation}
\frac{dp}{dt}=\left(2e^2\epsilon-4ne^2\sqrt{p^2+m^2} \right )^{1/2}.
\label{bose_clasic_eq}
\end{equation}
The solution of this equation is oscillatory with period
\begin{equation}
T=4\int_{0}^{p_{\rm max}} \frac{dp}
{\left ( 2e^2\epsilon-4ne^2\sqrt{p^2+m^2} \right )^{1/2}}\ .
\label{bose_clasic_T}
\end{equation}
Here $p_{\rm max}$ is defined by the zero of the denominator. In the case of a
strong initial electric  field,  the  particles  and  antiparticles  are
accelerated until they approach the velocity of light.  At  this  point  the
electric current stops increasing, and the electric field degrades until  it
changes  its  direction  and  accelerates  the  particles  in  the  opposite
direction.  If the initial
field is very intense, then most of the time the field is strong enough  to
keep
the particles flowing almost at the speed of light. In this case we are able
to estimate  the period $T$, since the particles are  ultrarelativistic.  In
this limit $p^2\gg m^2$ and therefore  (\ref{bose_clasic_T}) yields
\begin{equation}
T_{\it UR}=\frac{2\sqrt{nm+E^2_{\rm max}}}{ne}\ ,
\label{bose_Tur}
\end{equation}
where $E_{\rm max}$ is the amplitude of  the  electric  field.  The current
saturates at
\begin{equation}
\vert j^{\rm max}_{\it UR}\vert = 2en\vert v^{\rm max}\vert = 2en\ .
\label{bose_jur}
\end{equation}
If  the  initial
electric field is weak, the problem  becomes  nonrelativistic  and  the  set
(\ref{bose_classical}) is reduced to a simple  harmonic oscillator  equation
with a frequency
\begin{eqnarray}
\omega^2_p=\frac{2ne^2}{m}\,, \quad
T_{\it NR}=\frac{2\pi}{e}\sqrt{\frac{m}{2n}}\ .
\label{T_nr}
\end{eqnarray}
This is the well-known nonrelativistic plasma frequency. It is evident  that
for a fixed particle density $n$, the plasma  frequency
decreases as the electric field increases.
 
\begin{figure}
\begin{center}
\epsfig{file=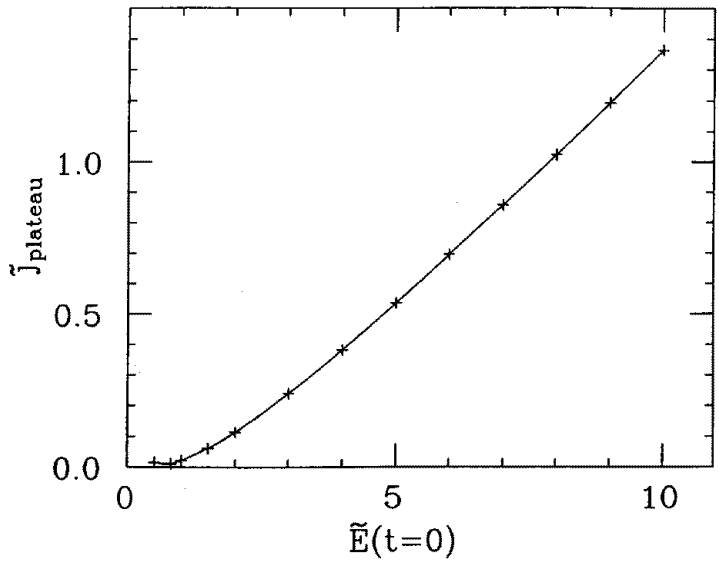,width=10cm}
\end{center}
\caption{Scaled  current  $\tilde  \jmath$  at  the  peak  of  the  first
oscillation, as a function of the initial scaled electric field $\tilde  E$.
\label{fig2_2}}
\end{figure}

The saturation of the first oscillation in our calculation is thus  easy  to
understand.  At  the  very  beginning  of  the  evolution,   particles   and
antiparticles are created and accelerated by the electric field. The current
$j$ saturates as $v$ is driven to the speed of light.  Eq.~(\ref{bose_jur})
gives us
a method for estimating the number of pairs $n$
created in  the  first  rise  of
$j$, allowing us to evade the ambiguities  in  the  definition  of  particle
number inherent in the adiabatic approach. We display a graph  of  the  peak
current in the first oscillation as a  function  of  the  initial  field  in
Fig.~\ref{fig2_2}.
 
In subsequent oscillations, $n$ is larger and the  peak  value  of
$E$  is  weaker because of particle  production,  so  the  particles  remain
nonrelativistic even at the peak of the current. At  late  times,  when  the
envelope of the electric field approaches a constant, that is, when  further
particle production is negligible, one can estimate the number of  particles
from (\ref{T_nr}).
 
\begin{figure}
\begin{center}
\epsfig{file=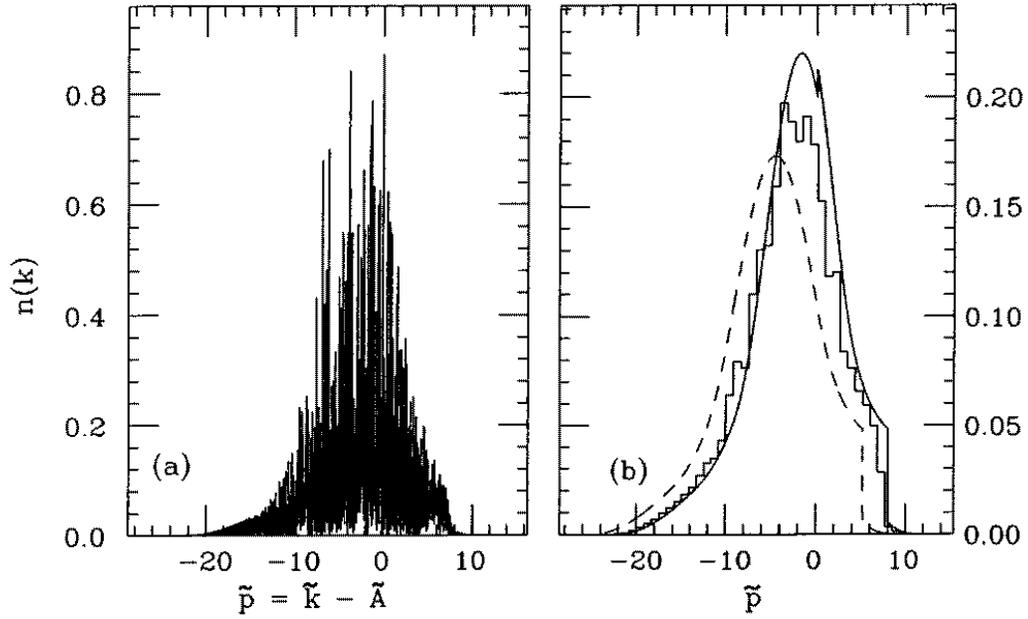,width=14cm}
\end{center}
\caption{(a) Momentum distribution of produced  pairs,  for  the  evolution
shown in Fig.~\ref{fig2_1}(a), at  time  $\tau=550.$   The  abscissa  is  the  scaled
kinetic momentum $\tilde p\equiv\tilde k-\tilde  A$,  with  $\tilde  k\equiv
k/m$.  (b)  Data  of  (a)  after  smoothing  (histogram),  compared  with
Boltzmann-Vlasov model.  The smooth curve is for the model
with Bose enhancement, the broken curve for the model without.
\label{fig2_3}}
\end{figure}

We can also evaluate the density of particles
by calculating the number operator at some late time, with the
expectation that at late times the number density is  almost  conserved.
We define the particle number density by expanding the exact  bosonic  field
operator $\Phi$ in terms of the zeroth-order mode functions
\begin{equation}
f_k^{(0)}(t)\equiv \frac{1}{\sqrt {2 \omega_{\bf k}(t)}}
\,\exp\left[- i\int^t\omega_k(t') dt'\right]\ .
\label{bose_fzero}
\end{equation}
Then the corresponding creation and annihilation
operators, $a_k^{(0)}$~and~$b_k^{(0)}$,
become time-dependent, and the vacuum expectation value of the
number operator
\begin{equation}
n(k;t)\equiv\langle 0\vert a^{(0)\dagger}_{k}(t) a^{(0)}_{k}(t) \vert 0
\rangle
\label{bose_n}
\end{equation}
may be computed by a Bogolyubov transformation from the basis functions
(\ref{PRL_7}) to the basis functions (\ref{bose_fzero}).
(See Appendix A for details.)
In the limit of large times we define
\begin{equation}
\lim_{t\to\infty} a_k^{(0)}(t)\equiv a_k^{\rm out}.
\label{bose_aout}
\end{equation}
The expectation value of the number operator for large times reads
\begin{eqnarray}
n(k)&=&\langle {0\vert a_k^{{\rm out}\dagger}a_k^{\rm out}
\vert 0}\rangle
\nonumber\\
&=&\lim_{t\to\infty}{1\over4\Omega_k\omega_k}
\left[(\Omega_k-\omega_k)^2+{1\over4}\left(
(\dot\Omega_k/\Omega_k) - (\dot\omega_k/\omega_k)\right)^2\right].
\label{bose_density}
\end{eqnarray}
We plot this quantity in Fig.~\ref{fig2_3}.
 
The rapid oscillations in $n(k)$ are related to rapid oscillations
in the integrand in (\ref{PRL_9}), which develop with time.
From (\ref{bose_addition}) we obtain
\begin{eqnarray}
\frac{1}{\Omega_k (t)}&=&\frac{1}{W_k (t)}\left[ \vert c_{1k}\vert ^2
+\vert c_{2k}\vert ^2
+ c_{1k}^{\ast}c_{2k}\exp{\left (2i\int^t W_k (t^{\prime})dt^{\prime}\right
 )}\right .
\nonumber \\
&&+ \left . c_{1k}c_{2k}^{\ast}\exp{\left (-2i\int^t W_k
 (t^{\prime})dt^{\prime}\right )} \right ].
\label{bose_oscillations}
\end{eqnarray}
Since  $W_k(t)\simeq  \omega_k(t)$ in the absence of interaction [see
the discussion  after  Eq.~(\ref{bose_Fulling_7_16})], it is evident from
(\ref{bose_oscillations}) that for a fixed time, $\Omega_k^{-1}$  oscillates
as a function of $k$.
At a fixed time $t$, this expression oscillates in momentum space with a
frequency on the order of $t$.
As time goes by, then, the frequency grows, and hence the integrands
in the current (\ref{PRL_9})  and
the particle distribution (\ref{bose_density}) also oscillate.
This kind of oscillation appears even if the initial pair-correlation
density (related to zitterbewegung \cite{Dutch}) is zero.
 
When   the   interaction   is    switched    off,
$\dot{\omega}_k=0$ but $\dot{\Omega}_k\neq 0$. It  is  clear  that  in  the
absence of interaction the expectation value of the  number  operator should
be  a constant.  One finds that the particle density in
(\ref{bose_density}) is indeed time-independent for $\dot\omega_k = 0$ by
calculating $\dot n(k)$ using (\ref{PRL_8}).

\subsection{Phenomenological kinetic approach}
 
The simple classical picture presented in the preceding subsection can explain
gross features of the time evolution of our system, such as the early
saturation of the current and, for that matter, the very existence of
plasma oscillations.
Particle creation is not included, nor is the distribution of
the created particles in momentum.
The inclusion of these features requires a more sophisticated model, and
a relativistic kinetic equation suggests itself.
The original Boltzmann-Vlasov equation, however, conserves particle number,
and thus one must add an explicit source term.
 
Many calculations in the realm  of  nucleus--nucleus  collisions  have  been
based  on  such  a   phenomenological   model [39--42].
In the case of a homogeneous electric field in 1+1 dimensions,
the relativistic kinetic equation is
\begin{equation}
\frac{\partial f}{\partial t}+eE\frac{\partial f}{\partial p} =
\frac{dN}{dt\,dx\,dp}\ ,
\label{PRL_11}
\end{equation}
where $f(p,t)$ is the ($x$-independent) classical phase-space  distribution,
expressed as  a  function  of  the  {\it  kinetic}  momentum  $p$,  and  the
right-hand side is the boson pair-production rate.
For the latter, one assumes the applicability of Schwinger's formula,
\begin{equation}
\frac{dN}{dt\,dx\,dp}={\vert  eE(t)\vert  }
\log \left[1+\exp \left(-\frac {\pi m^2}{\vert  eE(t) \vert }\right)\right]
\, \delta (p)\ ,
\label{PRL_11a}
\end{equation}
despite the fact that here the electric field is {\it not} constant in
time.
The factor $\delta(p)$ indicates the usual WKB assumption
that the particles are produced at rest \cite{CNN}.
Initially,    $f(p,0)=0$.
Eq.~(\ref{PRL_11}) may be solved using the  characteristics  ${dp}/{dt}=eE$,
giving
\begin {eqnarray} f(p,t)&=&\int_{0}^{t}dt'\,{\vert
eE(t')\vert } \,\log\left[1+\exp\left(-\frac{\pi m^2}{\vert eE(t')\vert
}\right)\right]\, \nonumber \\
&&\qquad\hbox{}\times\delta\bigl(p-eA(t')+eA(t)\bigr)\ .
\label{PRL_12a}
\end{eqnarray}
The $\delta$-function allows us to perform the integration, whence
\begin{eqnarray}
f(p,t)=\sum_{i} \log \left[1+\exp\left(-\frac{\pi m^2}
{\vert eE(t_i)\vert }
\right)\right],
\label{PRL_12b}
\end{eqnarray}
where the $t_i$'s fulfill $p+eA(t)-eA(t_i)=0$ and $t_i<t$.
 
The kinetic equation is coupled to the Maxwell equation,
\begin{equation}
\frac{d^2\!A}{dt^2}=j_{total}=j_{cond}+j_{pol}\ ,
\label{PRL_13}
\end{equation}
where the conduction current is
\begin{equation}
j_{cond}=2e\int\frac{dp}{2\pi}\,\frac{p}{\epsilon_p}\,f(p,t)\ ,
\label{PRL_14}
\end{equation}
with $\epsilon_p\equiv\sqrt {p^2  +  m^2}$,  and  the  polarization  current
is \cite{Gatoff87}
\begin{equation}
j_{pol}=\frac{2}{E}\int \frac{dp}{2\pi}\,\epsilon_p\frac{dN}{dt\,dx\,dp}\ .
\label{PRL_15}
\end{equation}
[The factors of 2 in  (\ref{PRL_14})  and  (\ref{PRL_15})  account  for  the
contributions of the \break antiparticles.] Inserting  (\ref{PRL_12a})  into
(\ref{PRL_13}) reduces the system to a single equation,
\begin{eqnarray}
\frac{d^2 \tilde{A}}{d\tau ^2}&=&\frac{e^2}{\pi m^2}\int_{0}^{\tau}\,d\tau'
\frac{\tilde{A}(\tau')-\tilde{A}(\tau)}{\sqrt{\left[\tilde{A}(\tau')-\tilde{A}
(\tau)\right]^2+1}}
\left\vert  \tilde{E}(\tau')\right\vert\nonumber\\
&&\qquad\mbox{}\times
\log \left[1+\exp \left(-\frac {\pi}{\left\vert \tilde{E}(\tau')
\right\vert }\right)\right] \nonumber \\
&& \mbox{}+\frac{e^2}{\pi m^2}\,{\rm sign}\left(\tilde{E}(\tau)\right)
\log \left[1+\exp\left(-\frac {\pi}{\left\vert \tilde{E}(\tau)\right\vert }
\right)\right],
\label{PRL_16}
\end{eqnarray}
in terms of the dimensionless variables $\tilde{A}\equiv eA/m$, $\tilde{E}$,
and $\tau$.
 
Equation~(\ref{PRL_11}), widely used in the literature, omits a statistical
factor
which should be present.
Roughly speaking, the source term should contain the Bose-Einstein factor
$\bigl(1+f(p,t)\bigr)\bigl(1+\bar f(-p,t)\bigr)$, where $\bar f$ is the
antiparticle density.
The $CP$ symmetry of the problem (and of the chosen initial conditions)
implies that $\bar f(-p,t)=f(p,t)$,
and thus the statistical factor is $\bigl(1+f(p,t)\bigr)^2$.
Detailed balance, however, dictates that there should be a loss term as
well, due to particle annihilation, containing the same matrix element
but the statistical factor $f(p,t)\bar f(-p,t)=\bigl(f(p,t)\bigr)^2$.
The difference of the gain and loss terms is thus proportional\footnote{
We thank J.-P.~Blaizot for this simple argument.  For a full
derivation of the statistical factor see \cite{PRD}.}
to
$\bigl(1+f(p,t)\bigr)^2-\bigl(f(p,t)\bigr)^2=1+2f(p,t)$.
We thus replace (\ref{PRL_11a})  with
\begin{equation}
\frac{dN}{dt\,dx\,dp}=\bigl(1+2f(p,t)\bigr) {\vert eE(t)\vert }
\log \left[1+\exp \left(-\frac{\pi m^2}{\vert eE(t)\vert }\right)\right]\,
\delta(p)\ .
\label{PRL_17}
\end{equation}
The modifications to (\ref{PRL_12a})--(\ref{PRL_12b}) and (\ref{PRL_16})
are obvious.
In (\ref{PRL_12b}), note carefully the strict inequality $t_i<t$.
 
\subsection{Comparison in 1+1 dimensions}
 
Let us compare the results of the phenomenological kinetic model
with those  of  our  semiclassical  analysis.  The  time
evolution of $\tilde{E}$ and $\tilde\jmath$ in the kinetic theory is   shown
in the dashed and dotted curves of Fig.~\ref{fig2_1}, where we see that
there is good  quantitative  agreement  between  the
results obtained with the two very different methods, as long as Bose-Einstein
enhancement is properly taken into account.\footnote{The enhancement is not
significant in the case of Fig.~\ref{fig2_1}(a), where the field is comparatively
weak and particle production is slow.}
Still, the  oscillations  are
faster and the electric fields  decay  more  rapidly  in  the quantum
calculation than in the Boltzmann-Vlasov model.
 
The  distribution  function  $f(p,t)$,   measured   after   several   plasma
oscillations, may be compared with the $n(k)$ of the  quantum  theory  after
the latter is smoothed, as shown in Fig.~\ref{fig2_3}.  Naturally,  the  curves  have
different normalizations and a relative displacement  due  to  the  slightly
different values of $j$ and $A$.
The kinetic theory is quite successful at reproducing the (smoothed) quantum
result.
One can thus use  the  kinetic-theory  model  to  explain
various features of the particle distribution, such as the sharp  edges  and
tails.
 
\begin{figure}
\begin{center}
\epsfig{file=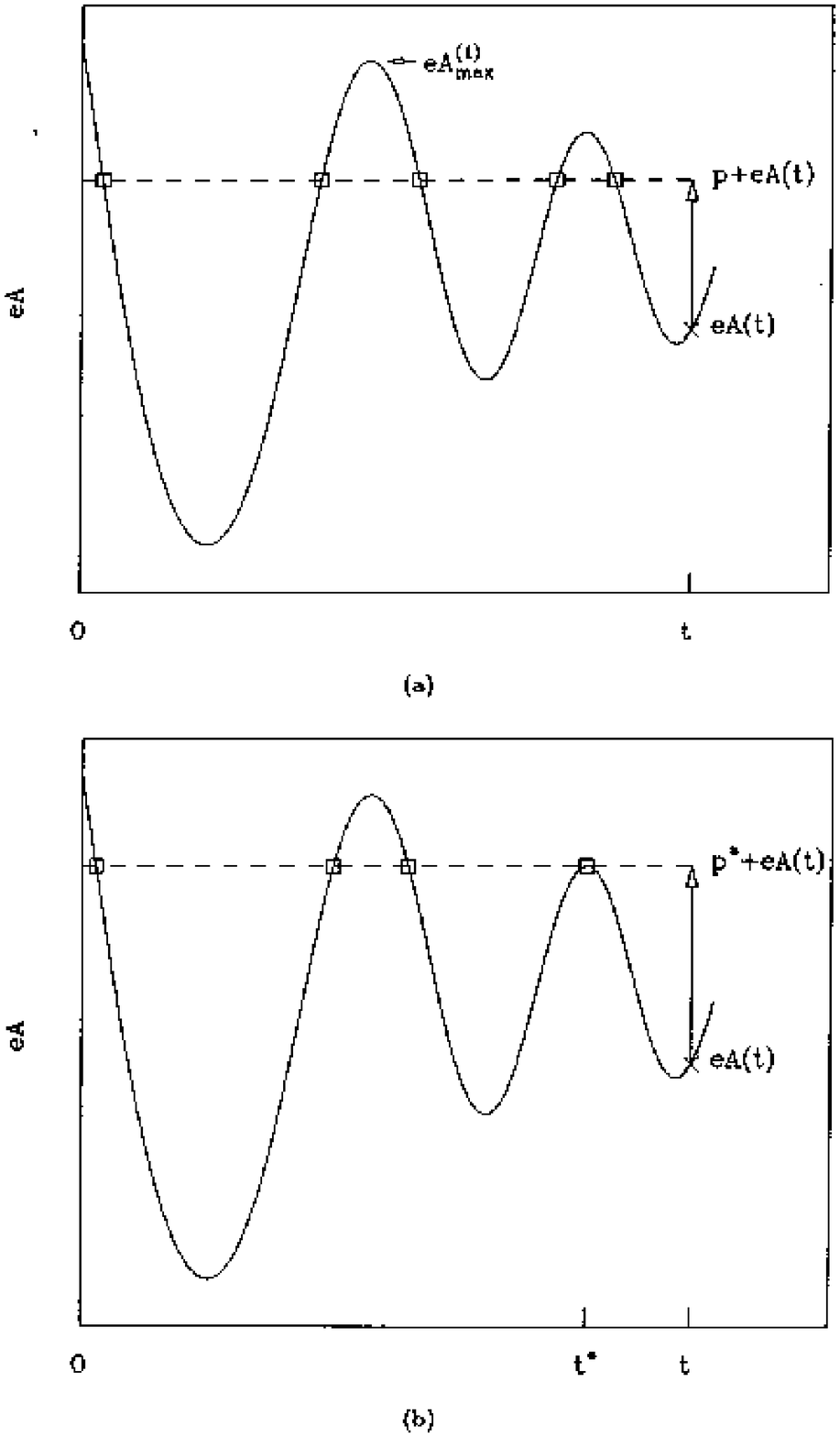,width=8.5cm}
\end{center}
\caption{(a) Schematic plot of $A(t)$, showing graphical construction
which gives $f(p,t)$ from (\ref{PRL_12b}).
The boxes represent the set of $t_i$ corresponding to a given value
of $p$, which is the length of the vertical arrow at $t$.
We define
$A^{(1)}_{\rm max}$ to be the value of $A(t)$ at the highest stationary point.
(b) Construction of $f(p,t)$ when $p=p^*$, a branch point where terms in
(\ref{PRL_12b}) disappear.
\label{fig2_4}}
\end{figure}

\begin{figure}
\begin{center}
\epsfig{file=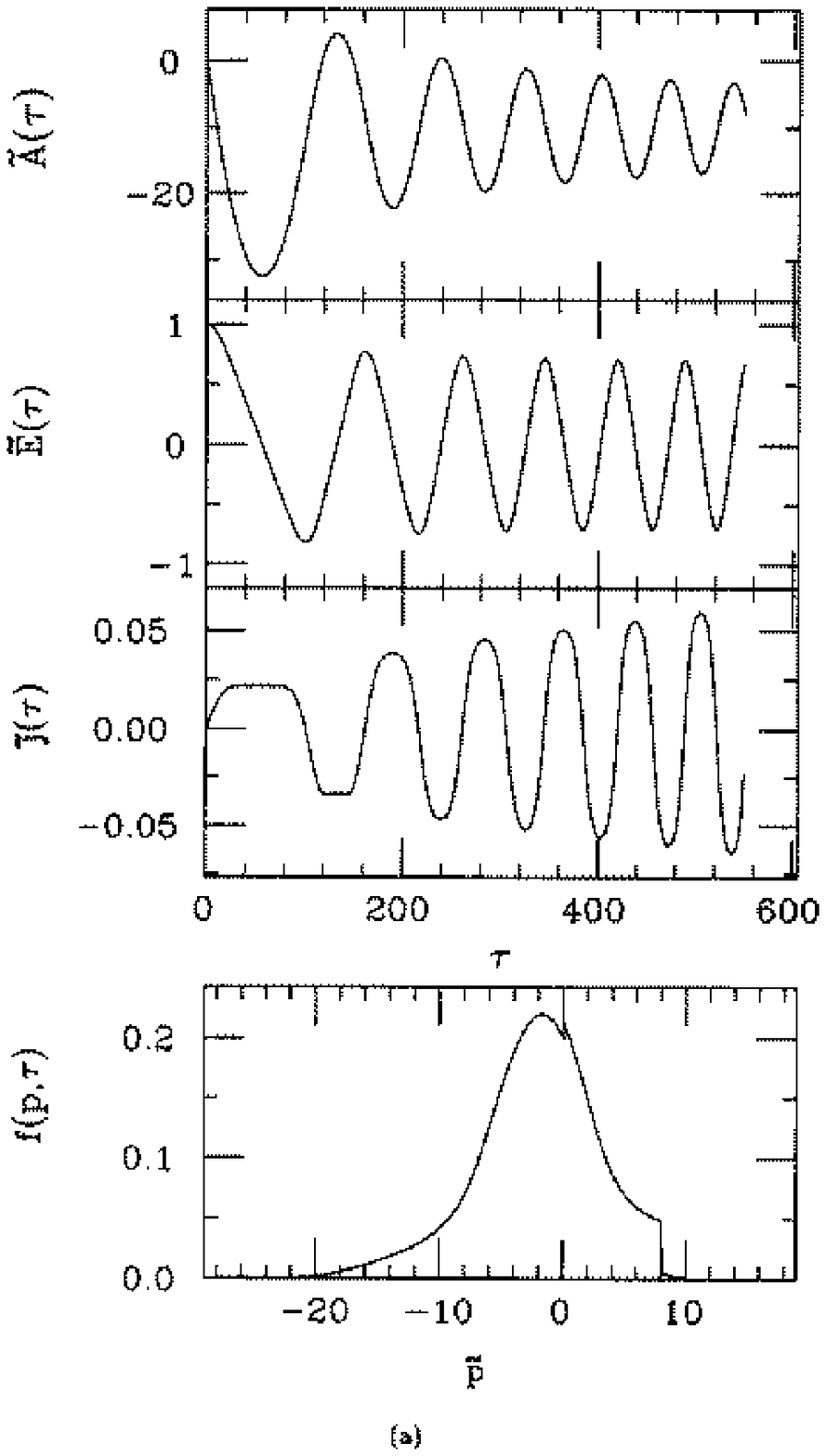,width=6.4cm}
\epsfig{file=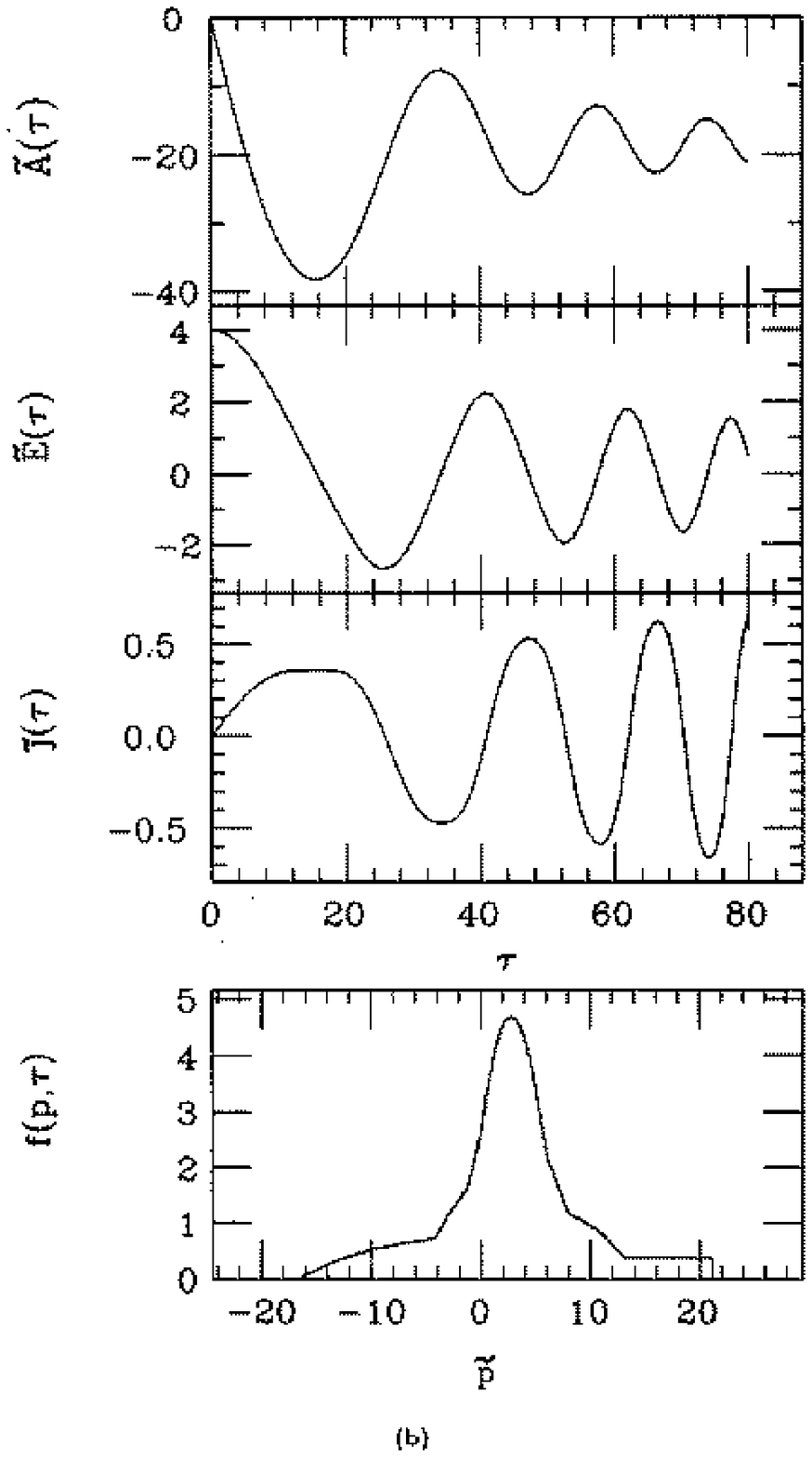,width=6.4cm}
\end{center}
\caption{(a) Time evolution of scaled  gauge  field  $\tilde  A$,  electric
field $\tilde E$, and current $\tilde \jmath$ for the  Boltzmann-Vlasov  model
with the stimulated pair creation correction.
The initial field is  $\tilde  E=1.0$ and the coupling is
$e^2/m^2=0.1$.
Also shown is the momentum distribution $f$ of produced pairs at time
$\tau=550$.
(b) Same as  (a),  but  for initial $\tilde E=4.0$, with $f$ measured
at $\tau=80$.
\label{fig2_5}}
\end{figure}

Let us investigate  the
formal solution given in  (\ref{PRL_12b}). The sum in  (\ref{PRL_12b})
consists of terms for which the following condition is fulfilled:
\begin{equation}
p+eA(t)-eA(t_i)=0, \quad t_i<t .
\label{bose_delta_argument}
\end{equation}
This is represented graphically in Fig.~\ref{fig2_4}.
Eq.~(\ref{bose_delta_argument}) shows
that the maximal and minimal values  of  $A$
at all times up to $t$ determine the  highest  and  lowest  momenta  in  the
momentum distribution function, since for $p>eA_{\rm max}-eA(t)$ and for
$p<eA_{\rm min}-eA(t)$ there is no solution for $t_i$.
Now consider Fig.~\ref{fig2_5}.
In the case where $\tilde{E}_{\tau=0}=4$ the
$\tilde{A}$ field is always negative, never crossing zero after $\tau=0$.
By examining Fig.~\ref{fig2_5}(b) we can explain the existence of the  plateau  on
the right-hand side of the distribution function $n(k)$: For momenta in the
range $eA^{(1)}_{\rm max}-eA(t)<p<eA(0)-eA(t)$,
where $A^{(1)}_{\rm max}$ is defined in Fig.~\ref{fig2_4},
the only contribution to the sum
(\ref{PRL_12b}) results from $t_i$ near zero, and the electric field at these
early times is changing slowly from its initial value.
The momentum distribution drops suddenly to zero at $p=eA(0)-eA(t)$, where
the only solution to (\ref{bose_delta_argument}) disappears and there are
thus no terms in the sum (\ref{PRL_12b}).
 
Consider now a generic point $\hat p$ in the momentum distribution at
time $t$.
There are  several  $t_i$'s  which  fulfill
condition (\ref{bose_delta_argument}) for $p=\hat p$.
If we increase $p$,  there
are branch points $p^*$ at which the number of $t_i$'s is reduced by two,  and
therefore there are two fewer terms in the sum (\ref{PRL_12b}).
For these values of $p$, the dashed line in Fig.~\ref{fig2_4}(b) just touches a local
extremum of $A$.
As $p\to p^*$, two of the times $t_i$ approach
$t^*$, the time at the extremum
of $A$.
Note, now, that $E(t^*)=-\dot A(t^*)=0$.
Terms  that  have   $E(t_i)=0$   do  not  contribute   to   the   sum
(\ref{PRL_12b}) and therefore
at  these branch points in $p$ the
distribution is continuous, though its derivative jumps.
 
There are, however, two special cases.
The point $p$ which corresponds to $t_i=0$ is one---here,
$E(t_i)$ is the initial value of the  electric  field,  which  is
{\em nonzero.}
The momentum distribution is thus always discontinuous at $p=eA(0)-eA(t)$.
The right edge of the distribution in Fig.~\ref{fig2_5}(b) is a special case of this;
in Fig.~\ref{fig2_5}(a), the discontinuity is at $\tilde p\simeq8$, in the interior of
the distribution.
 
The other special case is $p=0$.
Here we have the disappearance of the solution of (\ref{bose_delta_argument})
with $t_i=t$, at which, again, $E\not=0$.
Thus there is always a discontinuity at $p=0$.
 
In  the  case  of  $\tilde{E}(0)=1,$  shown  in Fig.~\ref{fig2_5}(a),
$A^{(1)}_{\rm max}>A(0)$, and the distribution function does  not
vanish at the point $p=eA(t_i=0)-eA(t)$.
Rather it decays smoothly to zero as
$p\to eA^{(1)}_{\rm max}-eA(t)$.
 
We offer a final observation regarding the Bose factor $1+2f$.
If the electric field were never to change sign, the factor would have
no effect, since new particles are only created at $p=0$, while the old
particles have been accelerated by the field.
Only if $E$ changes sign do some of the existing particles return to $p=0$
so that $f(p=0)$ is nonzero.  Thus the Bose factor is not relevant for
the usual case of the Schwinger mechanism in which the electric field is
fixed in time.
 
\subsection{Numerical results in 3+1 dimensions}
 
\subsubsection{Relation to the flux-tube model}
 
So far we have considered calculations in
$1+1$  dimensions.
Turning to 3+1 dimensions, let us begin by fixing the parameters and initial
conditions so as to correspond to the physical problem of particle creation
in the flux tube model \cite{CNN,BC86,BC88}.  The string
tension $\sigma$, interpreted as the energy per unit length  stored  in  the
flux  tube
between quark and antiquark, is given by
\begin{equation}
\sigma=\frac{1}{2}E^2 A\ ,
\label{bose_string}
\end{equation}
where $A$ is the cross-sectional area of the flux tube and  $E$ is the
strength of the
longitudinal electric field.  Applying Gauss' Law to the quark at either end
of the flux tube, one obtains
\begin{equation}
EA=e,
\label{bose_gauss}
\end{equation}
where $e$ is the effective charge of a quark (sweeping non-Abelian features
under the rug).
The  Regge  slope gives for
the string tension \cite{CNN}  $\sigma \simeq 0.2~{\rm GeV}^2$.
Taking $2.5~{\rm GeV}^{-1}$ for the radius of the tube,
and using constituent quark masses of
$0.35\ {\rm GeV}$,  one   finds  by  using
(\ref{bose_string}) and (\ref{bose_gauss}) that the  dimensionless field
strength in this elementary flux tube  is $\tilde{E}=eE/m^2 \simeq
3$, and the charge is $e \simeq 2.5$. The strength of the chromoelectric field
created in high-energy nuclear collisions should
be  stronger  than  this, since the sources will be in higher representations
of the color group \cite{Biro84,Kerman86};
in our calculations, therefore, we choose initial fields in the range
$\tilde E = $6--10, with $e^2 = $4--10.
 
\subsubsection{Numerical results}
 
We solve the system  (\ref{PRL_8}),  (\ref{CM_4_3}),  and  (\ref{CM_jR})  by
using Runge-Kutta along with
the iterative scheme described at the end of subsection
2.2.
Previous attempts to apply Runge-Kutta to the mode equation have encountered
instabilities \cite{Trafton71},
but we find that the combination
of Runge-Kutta with the iterative  scheme  proves  to  be  stable.
Simpson's rule is used  to  perform  the  integration  in  (\ref{CM_jR})  in
cylindrical coordinates.
 
\begin{figure}
\begin{center}
\epsfig{file=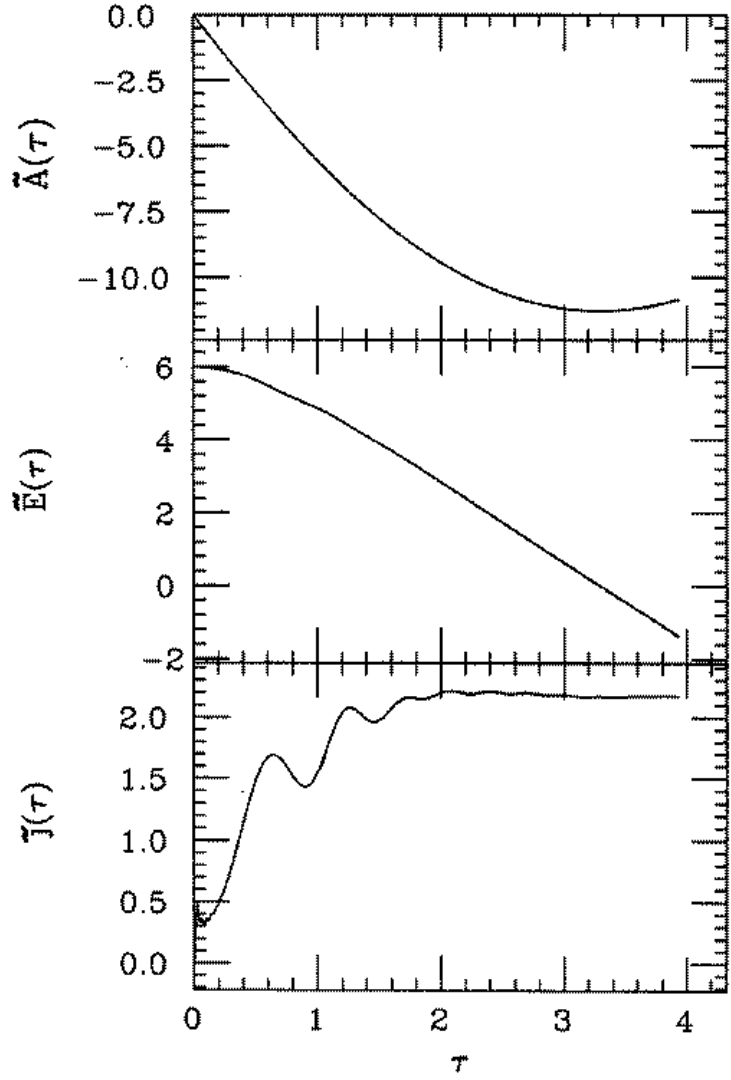,width=10cm}
\end{center}
\caption{Time evolution  of scaled gauge field $\tilde A$,  electric  field
$\tilde E$, and current $\tilde \jmath$ in $3+1$  dimensions,  with  initial
value $\tilde E=6.0$ and coupling $e^2=10$.
\label{fig2_6}}
\end{figure}

\begin{figure}
\begin{center}
\epsfig{file=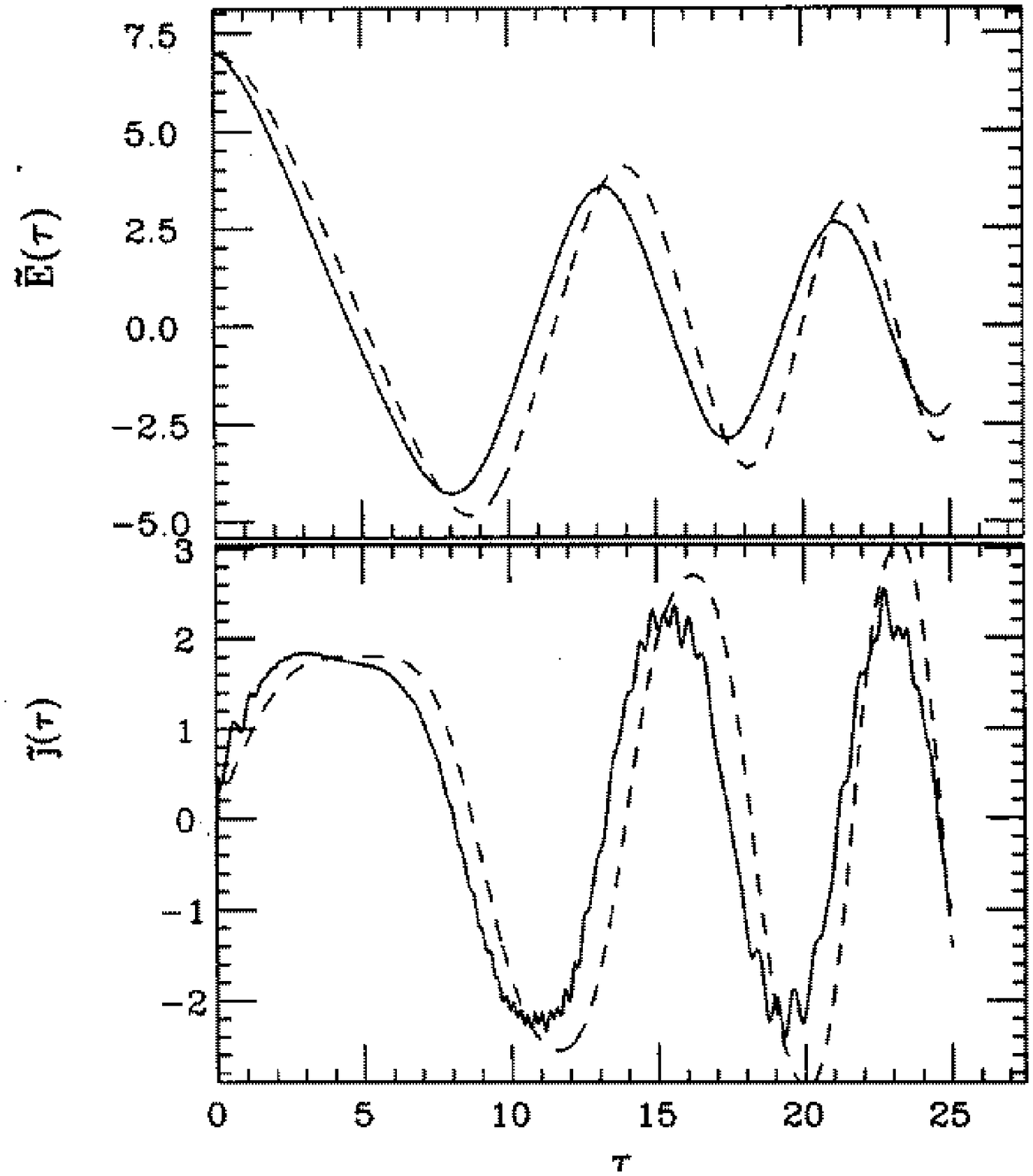,width=10cm}
\end{center}
\caption{
Time evolution  of scaled electric field $\tilde  E$  and  current
$\tilde \jmath$ in $3+1$ dimensions, with initial value $\tilde  E=7.0$  and
coupling $e^2=4$. Solid line is semiclassical scalar  QED,  dashed  line  is
the Boltzmann-Vlasov  model  with
the  stimulated pair-creation correction.
\label{fig2_7}}
\end{figure}

\begin{figure}
\begin{center}
\epsfig{file=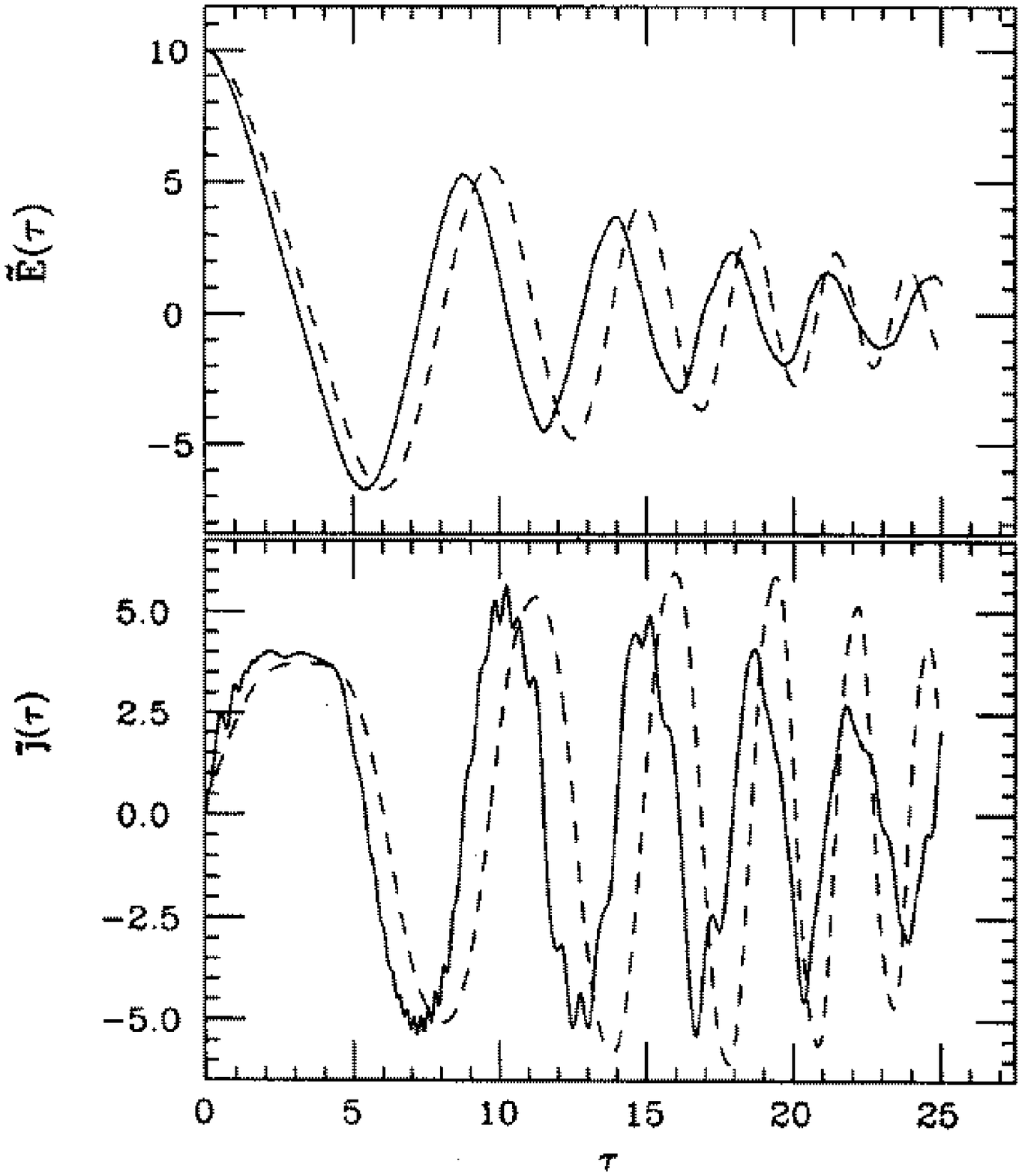,width=10cm}
\end{center}
\caption{
Same as Fig.~\ref{fig2_7}, but for initial $\tilde E=10.0$ and $e^2=5$.
\label{fig2_8}}
\end{figure}

In Fig.~\ref{fig2_6} we display the dimensionless  gauge  field
$\tilde{A}$, electric field $\tilde{E}$ , and current $\tilde\jmath$
as functions of the dimensionless time $\tau$.
We have set $\tilde{E}(0)=6$ and $e^2=10$,
and measure energy in units of $m$.
To get these high-precision results, we used a momentum matrix
grid of $200\times1000$  for  transverse
and  longitudinal  momenta, with  a  time  step
$d\tau=0.0005$ and momentum intervals  $dk_{\parallel}=dk_{\perp}=2\pi/25$.
One can see the saturation of the
current in the  first  oscillation, as in 1+1 dimensions.
To go to large times,  we
used smaller momentum grids and larger time steps. For the results
shown in Figs.~\ref{fig2_7} and~\ref{fig2_8} we
used momentum grids of $100\times400$ and $200\times200$, respectively,
and a time step of $d\tau=0.005$, with momentum intervals as in the case
shown in Fig.~\ref{fig2_6}.  Here we can  see plasma oscillations.
 
In order to compare the semiclassical calculation  to  the  phenomenological
model, we simply change $m^2$ to $m^2_{\perp}\equiv m^2+k_{\perp}^2$ in
(\ref{PRL_11})--(\ref{PRL_12b})   and   change   the   phase    space    in
(\ref{PRL_14})--(\ref{PRL_15}) to  $d^3{\bf p}/(2\pi)^3$.
The  result  in $3+1$ dimensions corresponding to (\ref{PRL_16}) is
\begin{eqnarray}
\frac{d^2 \tilde{A}}{d\tau^2}&=&\frac{e^2}{4\pi^3}\int_0^\tau\,d\tau'
\int d^2\tilde{p}_\perp
\left ( 2f(\tilde{p}_\parallel=0, \tilde{\bf p}_\perp,\tau')+1 \right
 )\nonumber \\
&&\qquad\mbox{}\times \frac{\tilde{A}(\tau')-\tilde{A}(\tau)}
{\sqrt{\left[\tilde{A}(\tau')-\tilde{A}
(\tau)\right]^2+\tilde{\bf p}_{\perp}^2 +1}}
\left\vert  \tilde{E}(\tau')\right\vert\nonumber\\
&&\qquad\mbox{}\times
\log \left[1+\exp \left(-\frac {\pi(1+\tilde{\bf p}_{\perp}^2)}{\left\vert
 \tilde{E}(\tau')
\right\vert }\right)\right] \nonumber \\
&& +\frac{e^2}{4\pi^3}\,{\rm sign}\left(\tilde{E}(\tau)\right)
\int d^2\tilde{p}_{\perp}
\sqrt{1+\tilde{\bf p}_{\perp}^2}
\left ( 2f(\tilde{p}_{\parallel}=0, \tilde{\bf p}_{\perp},\tau)+1 \right
 )\nonumber\\
&& \qquad\mbox{}\times
\log \left[1+\exp\left(-\frac {\pi(1+\tilde{\bf p}_{\perp}^2)}{\left\vert
 \tilde{E}(\tau)\right\vert }
\right)\right],
\label{bose_BV3D}
\end{eqnarray}
including Bose enhancement; the tilde represents
dimensionless  variables.  In  Figs.~\ref{fig2_7}  and~\ref{fig2_8}  it  is  seen  that   the
phenomenological Boltzmann-Vlasov  results  agree fairly well with  those  of
semiclassical QED.  Still, the quantitative agreement is not as good as in
1+1 dimensions; furthermore, the phenomenological model misses entirely the
rapid oscillations superposed on the plasma oscillation in $\tilde\jmath(\tau)$.
 
\section{Fermion pair production}
 
\subsection {Introduction}
We present here the extension of the semiclassical formalism
to spin-$\frac{1}{2}$ fields \cite{PRD}.
Again, we are limited to the case of a
spatially homogeneous, classical electric  field.
In  the preceding section we showed
that the results of our calculations   in   semiclassical
scalar
electrodynamics are very similar to those obtained from the
phenomenological relativistic Boltzmann-Vlasov equation. It is
of  interest to see if the same results obtain for fermions
which, unlike bosons, possess no  classical  limit,  and  for
which  Pauli  blocking, in place of Bose enhancement, implies
quite different consequences.
 
To apply adiabatic regularization to the fermion problem, we
shall
again need to express the Fourier amplitudes of the field
operators
in a WKB-like form.
This will allow us to isolate the ultraviolet divergences through
an
adiabatic expansion and to eliminate them through
renormalization.
The WKB parametrization will turn out to be a bit more complex
than
that for the bosons.
We shall further see that in order to make the current zero in
the
initial adiabatic vacuum, we shall need to express physical
expectation values by averaging over two different complete sets
of
linearly independent solutions of the Dirac equation.  Our
initial state is thus a {\it mixed} state.
In Subsection 3.2 we derive the coupled equations for the fields
in
the semiclassical limit of QED, and discuss the adiabatic
regularization procedure.
In Subsection 3.3 we present our numerical results in 1+1
dimensions
and compare them with the phenomenological Boltzmann-Vlasov
model.
 
\subsection {Fermion QED in the semiclassical mean field
approximation}
 
\subsubsection{Equations of motion}
 
The lagrangian density for electrodynamics is
\begin{equation}
{\cal L }=\bar {\Psi} i\gamma^{\mu}
(\partial_{\mu} +ieA_{\mu})\Psi -m\bar {\Psi}\Psi-
\frac {1}{4}F_{\mu \nu}F^{\mu \nu},
\label{PRD_2.1}
\end{equation}
where the metric convention is taken as $(+---)$.
We work explicitly in 3+1 dimensions, leaving the case of 1+1
dimensions for later.
For  the  $\gamma$-matrices we use the convention of Bjorken and
Drell \cite{BjD},
\begin{equation}
\begin{array}{lr}
\gamma^{0}=\left(\begin{array}{cc}
            I & 0\\
            0 & -I \end{array} \right)
 
            &\quad \gamma^{i}=\left(\begin{array}{cc}
             0 & \sigma^{i}\\
            -\sigma^{i} & 0 \end{array} \right),
 
\end{array}
\quad i=1,2,3
\label{PRD_2.2}
\end{equation}
where $I$ is the identity matrix and $\sigma^{i}$ are the Pauli
matrices.
 
Again we quantize only the matter field, while  the
electromagnetic field is treated  classically.
The coupled field equations read
\begin{eqnarray}
(i\gamma^{\mu}\partial_{\mu}-e\gamma^{\mu}A_{\mu}-m)\Psi (x)
= 0
\label{PRD_2.3}
\end{eqnarray}
and
\begin{eqnarray}
\partial_{\mu} F^{\mu \nu }=\langle j^{\nu} \rangle =
\frac{e}{2}\langle [\bar {\Psi},\gamma ^{\nu}\Psi]\rangle\ ,
\label{PRD_2.4}
\end{eqnarray}
where the expectation value is with respect to  the  initial
state  of  the spinor field. The commutator in  the  electric
current  guarantees  a  zero expectation value for  any
charge-conjugation  eigenstate.  Expressing  the solution of the
Dirac equation as
\begin{equation}
\Psi (x)=(i\gamma^{\mu}\partial_{\mu}-e\gamma^{\mu}A_{\mu}+m)\Phi
(x)\ , \label{PRD_2.5}
\end{equation}
and inserting (\ref{PRD_2.5}) into (\ref{PRD_2.3}), it follows
that $\Phi$ satisfies the quadratic Dirac equation \cite{GMF}
\begin{equation}
\left [ (i\partial_{\mu} -eA_{\mu})^2-\frac{e}{2}\sigma^{\mu
\nu}F_{\mu \nu} -m^2 \right ]\Phi (x)=0\ ,
\label{PRD_2.6}
\end{equation}
where $\Phi$ is a four-component spinor. Here we consider  the
case  where the electric field is spatially  homogeneous  so
that  the  field  strength $F^{\mu  \nu}$   depends  only  on
time.    Owing   to   homogeneity,   the Maxwell equations
(\ref{PRD_2.4})  allow  only  configurations
where $\langle j^0 \rangle =0$. We again take the electric field
to be in the direction of the $z$-axis and choose a gauge such
that only $A\equiv A^3 (t)$ is nonvanishing. Then the
second-order Dirac equation becomes
\begin{equation}
\left [ \Box +e^2 A^2 (t)+2iA(t)\partial_3-ie\partial_0
A(t)\gamma^0 \gamma^3 +m^2 \right ]\Phi (x)=0.
\label{PRD_2.7}
\end{equation}
 
Consider first the $c$-number Dirac equation corresponding to
(\ref{PRD_2.7}). Spatial homogeneity implies that there exist
solutions of the form
\begin{equation}
\phi_{{\bf k}s}(x) = e^{i{\bf k \cdot x}} f_{{\bf k}s}(t) \chi_s,
\label{PRD_2.8}
\end{equation}
where
\begin{equation}
\begin{array}{lr}
\chi_1=\left(\begin{array}{cc}
            1 \\
            0 \\
            1 \\
            0 \end{array} \right),
 
            & \chi_2=\left(\begin{array}{cc}
             0 \\
             1 \\
             0 \\
            -1 \end{array} \right),
\end{array}
\label{PRD_2.9}
\end{equation}
\begin{equation}
\begin{array}{lr}
 
              \chi_3=\left(\begin{array}{cc}
            1 \\
            0 \\
           -1 \\
            0 \end{array} \right),
 
            & \chi_4=\left(\begin{array}{cc}
             0 \\
             1 \\
             0 \\
             1 \end{array} \right).
 
\end{array}
\label{PRD_2.10}
\end{equation}
These spinors are chosen to be eigenvectors  of
$\alpha^3=\gamma^0\gamma^3$ in  the  representation
(\ref{PRD_2.2})  for  the  $\gamma$-matrices---the eigenvalues
of   $\alpha^3$   are   $\lambda_{s}=1$   for   $s=1,2$   and
$\lambda_{s}=-1$ for $s=3,4$. The spinors satisfy  the
normalization  and completeness conditions
\begin{eqnarray}
\sum_{\alpha =1}^{4} (\chi^{\dag}_r)^{\alpha}(\chi_s)_{\alpha}
&=& 2
\delta_{rs}\ ,
\nonumber \\
\sum_{r =1}^{4} (\chi^{\dag}_r)^{\alpha}(\chi_r)_{\beta} &=&
2 \delta^{\alpha}_{\ \beta}\ .
\label{PRD_2.11}
\end{eqnarray}
Substituting (\ref{PRD_2.8}) into (\ref{PRD_2.7}), it follows
that the  mode functions $f_{{\bf k}s}(t)$ satisfy
\begin{equation}
\frac{d^2 f_{{\bf k}s}(t)}{dt^2} +
\left [{\omega _{\bf k}^2 (t)-i\lambda_{s}e\frac{dA}{dt}}\right ]
f_{{\bf k}s}(t)=0\ ,
\label{PRD_2.12}
\end{equation}
with
\begin{equation}
\omega^2 _{\bf k} (t) \equiv p_{3}^{2} +{\bf
k}_{\perp}^2+m^2,\quad
{\bf k}_{\perp}^2\equiv k_1^2+k_2^2 ,\quad p_i \equiv
k^i-eA^i(t)\ .
\label{PRD_2.13}
\end{equation}
 
Equations (\ref{PRD_2.12})  are  second-order  differential
equations,  and therefore  for  each  $s$  there  are   two
independent   solutions.   Let $f^{+}_{{\bf k}s}$ and
$f^{-}_{{\bf k}s}$ be the two independent solutions of
(\ref{PRD_2.12}),  which   become   positive-   and
negative-frequency solutions in the absence of the electric
field. Clearly  at  the  moment  we have   eight   different
solutions   for   the    second-order    equation
(\ref{PRD_2.6}), namely, $f^{\pm}_{{\bf k}s}$ for $s=1,2,3,4$.
However,  the Dirac equation (\ref{PRD_2.3}) has only four
independent solutions.   If  we restrict ourselves to solutions
that belong to the set $s=1,2$  or  to  the set $s=3,4,$ we shall
see that from each set one  can  construct  a  linearly
independent set of solutions of the Dirac equation.
 
The form introduced in (\ref{PRD_2.8}) allows us to write $\psi$
as
\begin{equation}
\psi_{{\bf k}s}(x)\equiv \hat{D}\phi_{{\bf k}s} (x)
\equiv (i\gamma^{0}\partial_{0}+{\gamma}^i k_i
-e\gamma^{3} A_3+m)\phi_{{\bf k}s} (x)\ .
\label{PRD_2.14}
\end{equation}
Explicitly, the two sets of independent solutions of the Dirac
equation may be taken to be
 \begin{eqnarray}
\psi^{\pm}_{{\bf k}s}=
e^{i{\bf{k\cdot x}}}\hat{D}f^{\pm}_{{\bf k}s}\chi_s\ , \quad
s=1,2,
\label{PRD_2.15}
\end{eqnarray}
and
\begin{eqnarray}
\psi^{\pm}_{{\bf k}s}=
e^{i{\bf{k\cdot x}}}\hat{D}f^{\pm}_{{\bf k}s}\chi_s\ , \quad
s=3,4.
\label{PRD_2.16}
\end{eqnarray}
Using Eqs.~(\ref{PRD_2.14})--(\ref{PRD_2.16}) we find, for a
given
${\bf k},$
\begin{eqnarray}
{\psi^{\pm}_{r}}^{\dagger} {\psi^{\pm}_{s}} &=&
{\chi_{r}}^{\dagger}{f_{r}^{\ast \pm }}{\hat{D}}^{\dagger}
\hat{D}f_{s}^{\pm}\chi_{s}   \nonumber \\
&=&2 \left \{ \omega^2 _{\bf k} (t) {f_{r}^{\ast \pm }}
f_{s}^{\pm}+
{{\dot{f}}_{r}^{\ast \pm }}
\dot{f}_{s}^{\pm}\right.\nonumber\\
&&\quad\left.-i\lambda_{s}
p_3 \left [{f_{r}^{\ast \pm }}
\dot{f}_{s}^{\pm}-
{\dot{f}_{r}^{\ast \pm }} f_{s}^{\pm}\right ] \right
\}\delta_{rs}\ ,
\label{PRD_2.17}
\end{eqnarray}
where either $r,s =1,2$ or $r,s=3,4$. An exactly analogous
formula  may  be derived    for     ${\psi^{\pm}_{r}}^{\dagger}
{\psi^{\mp}_{s}}$.     By differentiating  these  expressions
with  respect  to  time  and  by  using Eq.~(\ref{PRD_2.12}), it
can be readily verified that these  inner  products are
time-independent. As $t \rightarrow -\infty$  there  is  no
interaction between the fermion field and  the electromagnetic
field, and we can  choose two independent plane-wave solutions
for Eq.~(\ref{PRD_2.12}),
\begin{eqnarray}
\lim_{t \rightarrow -\infty}f^{\pm}_{{\bf k}s}=
c_{s} e^{\mp i\omega_{\bf k}t},
\label{PRD_2.18}
\end{eqnarray}
where $c_s$ are constants.
Insertion   of    these    free    solutions    in    the
relation    for ${\psi^{\pm}_{r}}^{\dagger} {\psi^{\mp}_{s}}$
yields immediately
\begin{equation}
 {\psi^{\pm}_{r}}^{\dagger}{\psi^{\mp}_{s}}=0 \ .
\label{PRD_2.19}
\end{equation}
Since this result is time-independent, it is valid at any time,
and each set $\psi^{\pm}_{s}$,  with  $s=1,2$  or  $s=3,4$,  is
a   complete   set   of linearly-independent solutions  of  the
Dirac  equation.  Note  that  these complete systems are not
identical, and orthonormality conditions holds  for each set
separately.  In principle, we need only one of these sets in
order to expand the field operator $\Psi$ in terms of
single-particle  solutions. In order to ensure that, with our
initial  conditions,  the  Dirac  current vanishes at $t=0$, it
is  advantageous  to  use  {\it  both}  sets  in  our
calculations.
 
We now construct the quantized spinor field operator in the form
\begin{eqnarray}
\Psi (x)&=&\int [d{\bf k}]\sum_{s =1,2}[b_{s}({\bf k})
{\psi^{+}_{{\bf k}s}}
+d_{s}^{\dagger}({\bf{-k}}){\psi^{-}_{{\bf k}s}} ]
\nonumber \\
&= &
\int [d{\bf k}]\sum_{s =3,4}[b_{s}({\bf k})
{\psi^{+}_{{\bf k}s}}
+d_{s}^{\dagger}({\bf{-k}}){\psi^{-}_{{\bf k}s}} ]\ ,
\label{PRD_2.20}
\end{eqnarray}
where the two lines show the field expressed in terms of the two
alternative bases. Here $[d{\bf k}]={d^3{\bf k}}/{(2\pi)^3}$. The
fermion fields obey canonical      anticommutation
relations,       $\{       \Psi_{\alpha}
(t,{\bf{x}}),\Psi^{\dagger}      _{\beta}      (t,{\bf{y}})\}=
\delta^3 ({\bf{x}}-{\bf{y}})\delta_{\alpha \beta}$.  The
creation  and  annihilation operators of each set,  $r,s=1,2$  or
$r,s=3,4,$ obey the standard
anticommutation relations,
\begin{eqnarray}
 \{ b_{r}({\bf k}),b_{s}^{\dagger}({\bf{q}}) \}=
 \{ d_{r}({\bf k}),d_{s}^{\dagger}({\bf{q}}) \}=
(2\pi )^3\delta^3({\bf k}-{\bf{q}})\delta_{rs}\ ,
\label{PRD_2.21}
\end{eqnarray}
if we impose the condition
\begin{eqnarray}
 {\psi^{\pm}_{r}}^{\dagger}{\psi^{\pm}_{s}}= \delta_{rs}\ ,
\label{PRD_2.22}
\end{eqnarray}
to normalize the mode functions.
 
We now calculate the expectation value of the electric
current.
For the sake of simplicity we choose the initial state to be the
vacuum annihilated by $b_r({\bf k})$ and $d_r({\bf k})$.
Using the anticommutation relations (\ref{PRD_2.21}) we find
\begin{eqnarray}
\langle 0\vert j^3\vert 0 \rangle &=&\frac{e}{2}
\langle 0\vert [\bar{\Psi},\gamma^3\Psi]\vert 0\rangle
\nonumber\\
&=&\frac{e}{2}\int [d{\bf k}]\sum_{s=1,2}\left \{
-{\psi^+_{{\bf k}s}}^{\dagger}\gamma^0\gamma^3
{\psi^+_{{\bf k}s}}
+{\psi^-_{{\bf k}s}}^{\dagger}\gamma^0\gamma^3
{\psi^-_{{\bf k}s}} \right \}.
\label{PRD_2.23}
\end{eqnarray}
Alternatively,
\begin{eqnarray}
\langle 0\vert j^3\vert 0\rangle =\frac{e}{2}\int [d{\bf k}]
\sum_{s=3,4}\left \{-{\psi^+_{{\bf
k}s}}^{\dagger}\gamma^0\gamma^3
{\psi^+_{{\bf k}s}}
+{\psi^-_{{\bf k}s}}^{\dagger}\gamma^0\gamma^3
{\psi^-_{{\bf k}s}}\right  \}.
\label{PRD_2.24}
\end{eqnarray}
Averaging the two expressions,
\begin{eqnarray}
\langle 0\vert j^3\vert 0 \rangle =\frac{e}{4}\int [d{\bf k}]
\sum_{s=1}^{4}\left \{-{\psi^+_{{\bf
k}s}}^{\dagger}\gamma^0\gamma^3
{\psi^+_{{\bf k}s}}
+{\psi^-_{{\bf k}s}}^{\dagger}\gamma^0\gamma^3
{\psi^-_{{\bf k}s}}\right \}.
\label{PRD_2.25}
\end{eqnarray}
This form will be useful when we turn to the adiabatic
regularization
of the current.
The other  components  of  the  current  are  zero  since  the
electric field is in the $z$-direction. Using (\ref{PRD_2.14}) we
find \begin{equation}
\gamma ^{0}\gamma ^{3} {\psi^{\pm}_{{\bf k}s}}
= \lambda_{s} \left [-i\gamma^{0}\partial_{0}
+{\bf{\gamma}}^{\perp}
{\bf k}_{\perp}
-(k_3-eA_3)\gamma^{3}+m \right ]{\phi^{\pm}_{{\bf k}s}},
\label{PRD_2.26}
\end{equation}
and thus (\ref{PRD_2.17}) and (\ref{PRD_2.22}) give
\begin{eqnarray}
&&{\psi^{\pm}_{{\bf k}s}}^{\dagger}\gamma^0\gamma^3
{\psi^{\pm}_{{\bf k}s}}=
\chi_{s}^{\dagger}f_{{\bf k}s}^{\ast \pm }{\hat{D}}^{\dagger}
\gamma^0\gamma^3 \hat{D} f_{{\bf k}s}^{\pm }\chi_{s} =
 \nonumber \\
&=&2\lambda_{s} \left \{ (k^2_{\perp}+m^2-p_3{^2})
f_{{\bf k}s}^{\ast \pm }f_{{\bf k}s}^{\pm}-
\dot{f}_{{\bf k}s}^{\ast \pm }
\dot{f}_{{\bf k}s}^{\pm}+i \lambda_{s}
p_3(f_{{\bf k}s}^{\ast \pm }
\dot{f}_{{\bf k}s}^{\pm}-
\dot{f}_{{\bf k}s}^{\ast \pm } f_{{\bf k}s}^{\pm}) \right \}
\nonumber \\
&=&\lambda_{s}\left
[4(k^2_{\perp}+m^2)\vert f_{{\bf k}s}^{\pm}\vert ^2 -1 \right ],
\label{PRD_2.27}
\end{eqnarray}
where the index $s$ is not  summed  over.  Inserting
(\ref{PRD_2.27})  into (\ref{PRD_2.25}) yields
\begin{equation}
\langle 0\vert j^3\vert 0\rangle =e \sum_{s=1}^{4}\int
[d{\bf k}](k^2_{\perp}+m^2)
\lambda_{s} \left (\vert f_{{\bf k}s}^{-}\vert ^2 -
\vert f_{{\bf k}s}^{+}\vert ^2\right ).
\label{PRD_2.28}
\end{equation}
From (\ref{PRD_2.17}) and (\ref{PRD_2.22}) it can be shown \cite{GMF}
that
\begin{equation}
2(k^2_{\perp} +m^2)\left (\vert f_{{\bf k}s}^{+}\vert ^2 +
\vert f_{{\bf k}s}^{-}\vert ^2\right )=1 .
\label{PRD_2.29}
\end{equation}
Eq.~(\ref{PRD_2.29}) then gives the current as
\begin{equation}
\langle 0\vert j^3\vert 0\rangle
=-2e \sum_{s=1}^{4}\int [d{\bf k}](k^2_{\perp} +m^2)
\lambda_{s}\vert f_{{\bf k}s}^{+}\vert ^2 .
\label{PRD_2.30}
\end{equation}
Averaging the current expectation value over the two bases as in
(\ref{PRD_2.25}) means that we prepare the initial system in a
mixed state.
 
\subsubsection{Adiabatic regularization}
 
As in the boson case of the previous section, the difficulty in
solving  the coupled  semiclassical  equations   (\ref{PRD_2.3})
and   (\ref{PRD_2.4}) originates from the fact that the
expectation value of the current
(\ref{PRD_2.30}) diverges in the interacting theory.  This
infinity can,  as before, be removed by charge  renormalization.
In order to isolate  the  ultraviolet  behavior  of  the
current integrand by adiabatic  expansion,  we  need  to  express
the  mode equations (\ref{PRD_2.12}) in a WKB-like form. The
generic  problem  is  to find a suitable  parametrization  for
the  solution  of  the  differential equation $\ddot{u}(t) +
\epsilon(t) u(t) = 0$, where  in  the  present  case $\epsilon$
is the complex quantity in square brackets  in  (\ref{PRD_2.12}).
Such a parametrization is found in \cite{Waterman73}, namely,
\begin{eqnarray}
f_{{\bf k}s}^+ (t) = N_{{\bf k}s}
 \frac {1}{\sqrt{2\Omega_{{\bf k}s}}} \exp\left \{ \int_{0}^{t}
\left ( -i\Omega_{{\bf k}s} (t^{\prime})
-\lambda_{s}  \frac {e\dot{A}(t^{\prime})}{2\Omega_{{\bf k}s}
(t^{\prime})} \right )
dt^{\prime}\right \}  ,
\label{PRD_3.1}
\end{eqnarray}
where $N_{{\bf k}s}$ are normalization constants  and
$\Omega_{{\bf k}s}$ is a real generalized
fre\-quen\-cy.\footnote{
The second solution $f_{{\bf k}s}^-$  for
the mode equation can be found by using its  Wronskian.  When
choosing  the form (\ref{PRD_3.1}) for $f_{{\bf k}s}^+$, the
second  solution   does  not have  a  simple  form,  and  for
this  reason  we  expressed  the   current (\ref{PRD_2.30}) in
terms of the   positive-frequency  solutions  only.} By
substituting (\ref{PRD_3.1}) into (\ref{PRD_2.12}) we obtain the
WKB-like equation for $\Omega_{{\bf k}s}$,
\begin{eqnarray}
\Omega^2_{{\bf k}s}(t)=
-\frac{\ddot{\Omega}_{{\bf k}s}}{2\Omega_{{\bf k}s}}
+\frac{3}{4}
\frac{\dot{\Omega}^2_{{\bf k}s}}{ \Omega^2_{{\bf k}s}}
+\omega^2_{\bf k}+\left( \frac{e\dot{A}}{2\Omega_{{\bf
k}s}}\right )^2 -\lambda_{s} \frac{e\ddot{A}}{ 2\Omega_{{\bf
k}s}}
+\lambda_{s} \frac{e\dot{A}\dot{\Omega}_{{\bf k}s}}{
\Omega^2_{{\bf k}s}}  . \label{PRD_3.2}
\end{eqnarray}
As in the boson case, the equation for $\Omega$ is a second-order
nonlinear differential equation.
 
This equation enables  us  to  study  the  large-momentum
behavior  of  the solutions.  As above, an adiabatic expansion
of  (\ref{PRD_3.2})
to second order  is needed to identify the  divergences  in  the
current  (\ref{PRD_2.30}).
Noting that
\begin{equation}
\ddot{\omega}_{\bf k}=\frac {-e\ddot{A}p_3}{\omega_{\bf k}}
+O(1/ \omega_{\bf k}) ,
\label{PRD_3.3}
\end{equation}
we have up to second order [i.e., iterating (\ref{PRD_3.2}) once]
\begin{equation}
\Omega_{{\bf k}s}=
\omega_{\bf k}-e\ddot{A}\left (\lambda_{s} \omega_{\bf k}
- p_3 \right ) /4\omega^3_{\bf k}
+O(1/ \omega^3_{\bf k}).
\label{PRD_3.4}
\end{equation}
Using the ansatz (\ref{PRD_3.1}) the current reads
\begin{eqnarray}
\langle 0\vert j^3\vert 0 \rangle
= -2e \sum_{s=1}^{4}\int [d{\bf k}](k^{2}_{\perp}+m^2)
\lambda_{s} \left [  \frac{\vert N_{{\bf k}s}\vert
^2}{2\Omega_{{\bf k}s}} \exp\left \{-\lambda_{s}\int \frac
{e\dot{A}(t^{\prime})}
{\Omega_{{\bf k}s}(t^{\prime})}dt^{\prime}\right \} \right].
\nonumber \\ \label{PRD_3.6}
\end{eqnarray}
Equations~(\ref{PRD_2.17}) and (\ref{PRD_2.22})
determine the normalization
constants, and it turns out that the expression in square
brackets is
\begin{eqnarray}
\frac{\vert N_{{\bf k}s}\vert ^2}
{\Omega_{{\bf k}s}}
 \exp\left \{ -\lambda_{s} \int \frac{e\dot{A}(t^{\prime})}
{\Omega_{{\bf k}s}(t^{\prime})} dt^{\prime}\right \}
&=&\left [
\omega^2_{\bf k}+\left ( \frac{-\lambda_{s} e\dot{A}}{
2\Omega_{{\bf k}s}}- \frac{\dot{\Omega}_{{\bf k}s}}{2\Omega_{{\bf
k}s}}\right ) ^2
+\Omega^2_{{\bf k}s}\right. \nonumber\\
&&\mbox{}-\lambda_{s} 2p_3\Omega_{{\bf k}s}\Biggr] ^{-1}
\nonumber \\
&\equiv &\Gamma_{{s}}({\bf k})\ .
\label{PRD_3.7}
\end{eqnarray}
With the identity
\begin{equation}
k^2_{\perp}+m^2=[\omega_{\bf k}+p_3][\omega_{\bf k}-p_3],
\label{PRD_3.8}
\end{equation}
we obtain
\begin{eqnarray}
\langle 0\vert j^3\vert 0\rangle &=&e\int {[d{\bf k}]}
\left \{  \frac{\omega_{\bf k}-p_3}
{\omega_{\bf k}+e\ddot{A}(\omega_{\bf k}+p_3)
/ 4\omega^3_{\bf k}}
 -  \frac{\omega_{\bf k}+p_3}
{\omega_{\bf k}-e\ddot{A}(\omega_{\bf k}- p_3)
/ 4\omega^3_{\bf k}}\right.
 \nonumber \\
&& + \,O(\omega^{-4}_{\bf k})\biggr\}  .
\label{PRD_3.9}
\end{eqnarray}
At large momentum, we approximate
\begin{eqnarray}
\frac {1}{1\pm e\ddot{A}[\omega_{\bf k} \pm  p_3]
/4\omega^4_{\bf k}}
\simeq
1\mp \frac{e\ddot{A}}{4\omega^4_{\bf k}}[\omega_{\bf k}\pm p_3]\
.
\label{PRD_3.10}
\end{eqnarray}
 After we perform the angular integrations and drop
terms that are odd functions of $p_3$, the Maxwell
equation becomes
\begin{eqnarray}
\ddot{A}&=&\langle 0\vert j^3\vert 0\rangle  ={-e^2\ddot{A}}
\int[d{\bf k}] \left ( \frac{1}{2\omega^3_{\bf k}} -
\frac{p_3^2}{2\omega^5_{\bf k}}\right)
 + ({\rm finite\; part}) \nonumber\\
&=& {-e^2\ddot{A}}\,\delta e^2 + ({\rm finite\; part}),
\label{PRD_3.11}
\end{eqnarray}
where
\begin{equation}
\delta e^2 \equiv \int [d{\bf k}]
\left ( \frac{1}{2\omega^3_{\bf k}} - \frac{p_3^2}{2\omega^5_{\bf
k}}\right) =\frac{1}{4\pi^2} \int_{0}^{\infty}dk\left[
\frac{k^2}{(k^2+m^2)^{\frac{3}{2}}}
- \frac{k^4}{3(k^2+m^2)^{\frac{5}{2}} } \right ].
\label{PRD_3.12}
\end{equation}
 
The current in (\ref{PRD_3.11}) diverges logarithmically, with
the  same divergence as the vacuum polarization $\Pi (q^2=0)$.
Renormalizing as usual, we define
\begin{equation}
e^2_{R}=e^2(1+e^2\delta e^2)^{-1}\equiv Ze^2 ,\quad \quad
A_R=Z^{-1/2}A  , \label{PRD_3.13}
\end{equation}
so that $eA=e_RA_R$. We can also write $Z=(1-e^2_R\delta e^2)$.
Multiplying (\ref{PRD_3.6}) by $Ze/e_R$  we obtain
\begin{eqnarray}
\ddot{A}_R-e^2_R\ddot{A}_R\delta e^2=e_R\sum_{s=1}^{4}\int [d{\bf
k}]
 (k^2_{\perp}+m^2)
(-\lambda_{s})\Gamma_{s}({\bf k}) \ .
\label{PRD_3.14}
\end{eqnarray}
Using (\ref{PRD_3.12}) and rearranging, we have
\begin{equation}
\ddot{A}_R=e_R\int [d{\bf k}]\left [(k^2_{\perp}+m^2)
 \sum_{s=1}^{4}(-\lambda_{s})\Gamma_{s}({\bf k})
+{e_R^2\ddot{A_R}} \left ( \frac{1}{2\omega^3_{\bf k}} -
\frac{p_3^2}{2\omega^5_{\bf k}}\right) \right],
\label{PRD_3.15}
\end{equation}
where the right-hand side is finite.
We drop the $R$ subscripts now.
 
As for the boson problem, we can define a remainder which is the
difference between the integrand in (\ref{PRD_3.6}) and its
adiabatic approximation. Examining  (\ref{PRD_3.9}),
(\ref{PRD_3.10}),
and (\ref{PRD_3.14}), we can write
\begin{eqnarray}
(k^2_{\perp}+m^2)  \sum_{s=1}^{4}(-\lambda_{s}) \Gamma_{s}({\bf
k})
=-\frac{2p_3}{\omega_{\bf k}}-\frac{e\ddot{A}}{2\omega^5_{\bf k}}
(\omega^2_{\bf k}-\pi^2_3) + R_{\bf k}(t)\ .
\label{PRD_3.16}
\end{eqnarray}
At  large  momentum  this  minimal
adiabatic approximation matches the exact integrand up  to  terms
that fall off as $O(1/\omega^3_{\bf k})$, so the remainder
$R_{\bf k}(t)$ falls off faster. Upon substituting
(\ref{PRD_3.16}) into  (\ref{PRD_3.15})  and using
(\ref{PRD_3.9})--(\ref{PRD_3.11}), the finite Maxwell equation
takes the form \begin{equation}
\ddot{A}=e\int [d{\bf k}]R_{\bf k}(t)  .
\label{PRD_3.17}
\end{equation}
Again, the subsidiary condition
(\ref{PRD_3.16})   defining   $R_{\bf k}$   is   an   intrinsic
part   of (\ref{PRD_3.17}).
 
As in the boson case, the initial conditions we impose are
\begin{equation}
\dot{A}(t=0) = -E_0\,,\quad  A(t=0)=0\,,
\label{PRD_3.18}
\end{equation}
and
\begin{equation}
\Omega _{{\bf k}s}(t=0) = \omega_{{\bf k}}(t=0)\,, \quad
\dot{\Omega} _{{\bf k}s}(t=0) = \dot{\omega}_{{\bf k}}(t=0)\,,
\label{PRD_3.19}
\end{equation}
which specifies the adiabatic vacuum.
Nonvacuum initial conditions may be handled in a  manner
analogous  to  the bosonic case by adding nonzero particle number
densities to the current expectation value.
 
Here, too, the choice of initial conditions is constrained by
demanding the consistency of the adiabatic expansion
(\ref{PRD_3.16}) with the Maxwell equation (\ref{PRD_3.17}).
By    substituting
(\ref{PRD_3.19})  into (\ref{PRD_3.7}) and  (\ref{PRD_3.16})  we
find  that $\ddot{A}(0) = 0$, but
\begin{equation}
R_{\bf k}(0)=\frac{ 2p_3 }{\omega_{\bf k} }
\left \{ 1 - \frac{1+e^2E^2_0/4\omega^4_{\bf k}}
{ \left[ (1+e^2E^2_0/ 8\omega^4_{\bf k})^2- (e^2E^2_0p_3
/ 8\omega^5_{\bf k})^2
\right ] } \right \}
\label{PRD_3.20}
\end{equation}
is not zero.
It turns out, however, that the integration over $\bf k$ in
(\ref{PRD_3.17}) gives
zero by charge conjugation symmetry\footnote{
The summation in (\ref{PRD_3.16}) over $s$, along with the
$\lambda_s$ factor, is essential for making the integrand odd in
$p_3$.
It is here that our adoption of a mixed initial state proves
its value.  Choosing only the $s=1,2$ basis,
for example, would have made it impossible to have a zero initial
current.} $p_3 \rightarrow -p_3$.
Thus the
initial conditions (\ref{PRD_3.18}) and (\ref{PRD_3.19}) are
consistent with the requirements of renormalization.
We emphasize that choosing $R_{\bf k}(0)=0$ is {\it not}
consistent with the initial conditions (\ref{PRD_3.19}).
 
Given such a set  of  consistent  initial  conditions,  we  can
solve the back-reaction    equations    (\ref{PRD_3.2}),
(\ref{PRD_3.16}),     and (\ref{PRD_3.17}) exactly as we did in
the scalar case. To do  this  we  take $R_{\bf k}$ to be zero at
an extremely large momentum as a trial value,  so that
$\ddot{A}$  can  be  extracted  from  (\ref{PRD_3.16}).   Now
having $\ddot{A}$,  we  use  (\ref{PRD_3.16})  to  extract
$R_{\bf k}$  for  each ${\bf k}$ up to this very large momentum.
Then,  substituting  $R_{\bf k}$ in (\ref{PRD_3.17}) we get  a
new  corrected  value  for  $\ddot A$.  This procedure may be
iterated until convergence for $\ddot A$ and
$R_{\bf k}$ is reached, after which one may proceed to the next
time step.
 
\subsection{Calculation in 1+1 dimensions}
 
\subsubsection{Dynamical equations}
 
For the fermion problem, we present numerical results only for
the
(1+1)-dimensional case.
Let  us  indicate  the
modifications needed in the equations.  The $\gamma$-matrices are
given by \begin{equation}
\begin{array}{lr}
\gamma^{0}=\left(\begin{array}{cc}
            1 & 0\\
            0 & -1 \end{array} \right) ,
 
            & \gamma^{1}=\left(\begin{array}{cc}
             0 & 1\\
            -1 & 0 \end{array} \right),
 
\end{array}
\label{PRD_4.1}
\end{equation}
and  $\gamma^1$ plays the role that $\gamma^3$ played in the
(3+1)-dimensional case.  In $A^0=0$ gauge, we
define $A^1=A(t)$, and the  second-order  Dirac
equation is
\begin{equation}
\left [ \Box +e^2 A^2 (t)+2iA(t)\partial_1-ie\partial_0
A(t)\gamma^0 \gamma^1 +m^2 \right ]\phi (x)=0.
\label{PRD_4.2}
\end{equation}
The Dirac equation in two dimensions has two independent
solutions:  Either \begin{equation}
 f_{k1}^+ \chi_1\,, \quad
 f_{k1}^- \chi_1 \quad
\label{PRD_4.3}
\end{equation}
or
\begin{equation}
 f_{k2}^+ \chi_2\,, \quad
 f_{k2}^- \chi_2
\label{PRD_4.4}
\end{equation}
may be taken as the basis  set  of  independent  solutions.  Here
$\gamma^0 \gamma^1   \chi_{{s}}=\lambda_{s}   \chi_{{s}}$   with
$\lambda_1=1$   and $\lambda_2=-1$, and
the spinors $\chi_{{s}}$ are given by
\begin{equation}
\begin{array}{lr}
\chi_1=\left(\begin{array}{cc}
            1 \\
            1 \end{array} \right),
 
            & \chi_2=\left(\begin{array}{cc}
             1 \\
            -1 \end{array} \right).
\end{array}
\label{PRD_4.6}
\end{equation}
We define
\begin{equation}
\omega^2 _k (t) \equiv p^{2} +m^2,\quad
p \equiv k-eA^1(t)\ .
\label{PRD_4.5}
\end{equation}
The Maxwell equation is
\begin{equation}
\ddot{A}=\langle 0| j^1| 0\rangle =2e\int \frac{dk}{2\pi}m^2
\{ | f_{k2}^{+}| ^2 - | f_{k1}^{+}| ^2   \}.
\label{PRD_4.7}
\end{equation}
In two dimensions there is no spin, and thus there are half as
many terms as in (\ref{PRD_2.30}) after summation.
 
Renormalization of (\ref{PRD_4.7})
can  be  done  in  the  same  way  as  in  subsection  3.3.  In
two-dimensional QED the charge renormalization is finite and  we
find  that $\delta  e^2=(6\pi  m^2)^{-1}$.  Therefore,  in  the
renormalized  Maxwell equation, $\ddot{A}$ can be isolated [in
contrast to (\ref{PRD_3.15})],  and we obtain
\begin{eqnarray}
\ddot{A}_R=\frac{e_R}{1-e^2_R \delta e^2} \int  \frac{dk}{2\pi}
\left \{m^2\sum_{s=1,2}\left(-\lambda_{s} \Gamma_{s}(k)\right )
+2\frac{p}{\omega_k} \right \}
\label{PRD_4.8}
\end{eqnarray}
where
\begin{eqnarray}
\Gamma_{{s}}(k)  \equiv
\left [\omega^2_{k}+\left ( \frac{-\lambda_{s} e_R\dot{A}_R}
{ 2\Omega_{ks}}-
\frac{\dot{\Omega}_{ks}}{2\Omega_{ks}}\right ) ^2
+\Omega^2_{ks}
-\lambda_{s} 2p \Omega_{ks}    \right] ^{-1}.
\label{PRD_4.9}
\end{eqnarray}
The last term in the braces in (\ref{PRD_4.8}) does not
contribute  to  the integral but was included for  numerical
purposes.  The  set  of  equations (\ref{PRD_3.2})   and
(\ref{PRD_4.8})    with   the   initial   conditions
(\ref{PRD_3.18}) and (\ref{PRD_3.20}) defines the  numerical
back-reaction problem.
 
\subsubsection{Numerical results}
 
\begin{figure}
\begin{center}
\epsfig{file=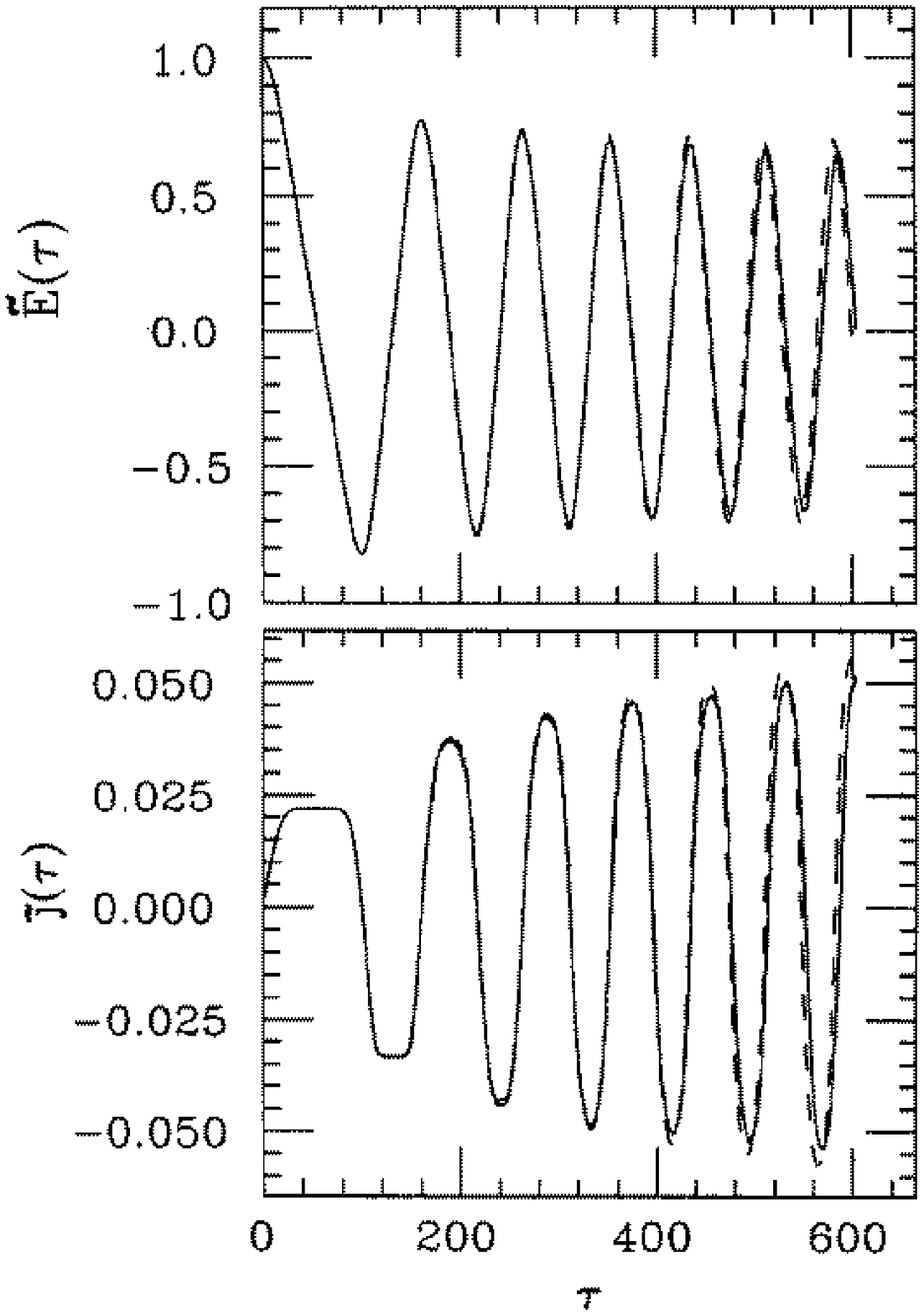,width=10cm}
\end{center}
\caption{Fermion pair creation in 1+1 dimensions:
Time evolution of the scaled electric  field  $\tilde  E$  and
current $\tilde \jmath$, with initial  value  $\tilde  E=1.0$  and  coupling
$e^2/m^2=0.1$.  Solid  line  is  semiclassical  QED,  and  dashed  line   is
Boltzmann-Vlasov model.
\label{fig3_1}}
\end{figure}

\begin{figure}
\begin{center}
\epsfig{file=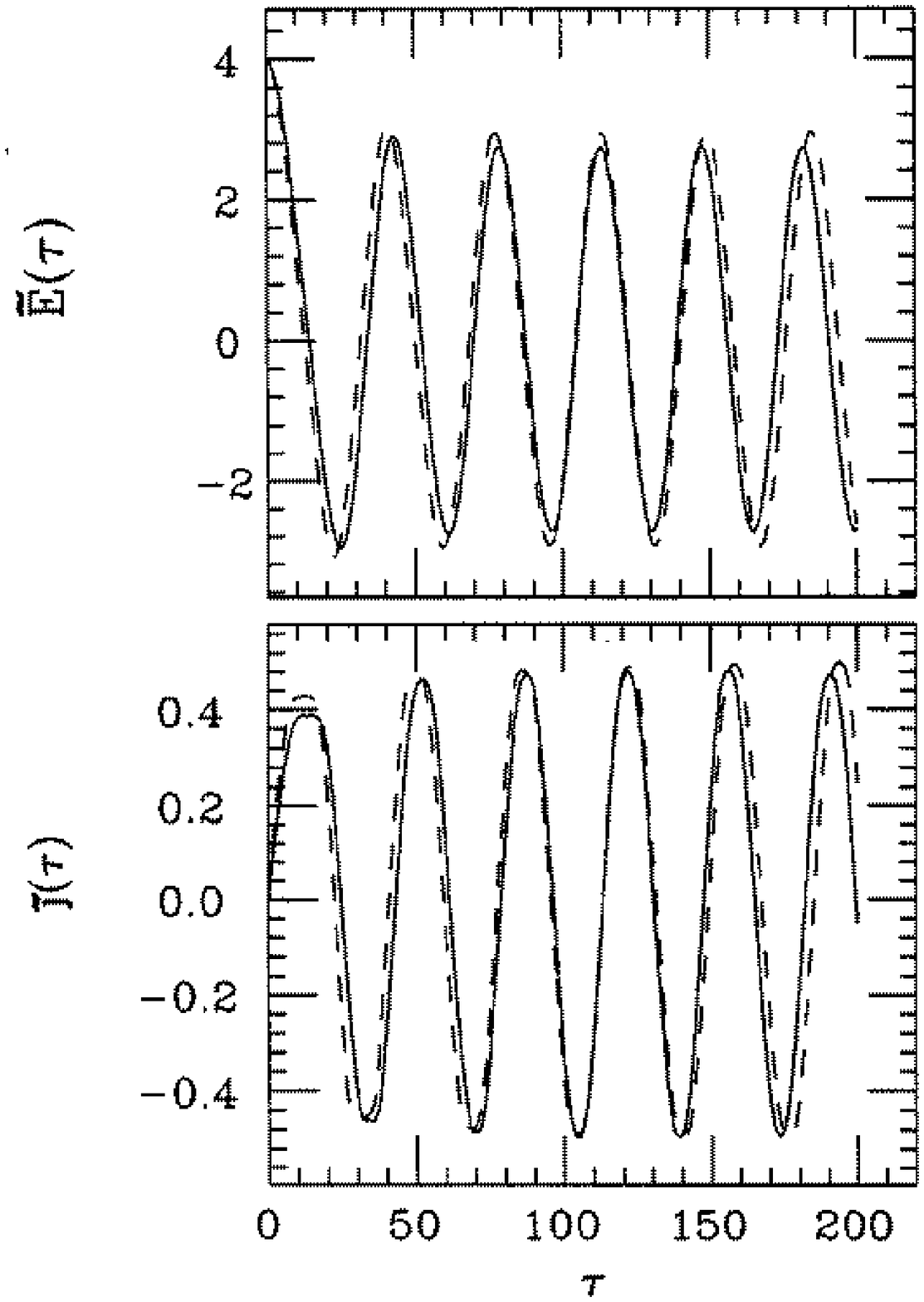,width=10cm}
\end{center}
\caption{As in Fig.~\ref{fig3_1}, but for initial field $\tilde E=4.0$.
\label{fig3_2}}
\end{figure}

We show in Figs.~\ref{fig3_1} and~\ref{fig3_2}  the time  evolution  of  the
scaled  electric field $\tilde E\equiv eE/m^2$  and  the  induced
current  $\tilde\jmath\equiv ej/m^3$ as functions of $\tau\equiv
mt$.  Here we used a time step of
$d\tau=5\times 10^{-4}$ and a  momentum  grid  with
$d\tilde{k}\equiv  dk/m =0.003$.   The results are similar to
those for bosons, with initial
particle creation followed by plasma oscillations.  In this case
the current saturates {\it twice} in the early stages of
evolution, each time because the particle velocity approaches
$c$, with more particles present (and hence a larger current) in
the second instance.
 
The amplitude of the electric-field oscillations
decreases substantially only in
the first few oscillations and remains almost constant at later
times. This means that essentially all of the
pair production happens in  the  first oscillations.
One sees that
in successive oscillations $n$ is larger and $E$ is  weaker,  and
therefore  the  frequency  of  oscillations  increases.
As we have noted,
the  frequency  of relativistic plasma  oscillations
depends not only on $n$ but also on the amplitude  of  $E$:
The  weaker  the
field the higher the frequency, in  contrast  to  the
nonrelativistic  case where the plasma frequency does not depend
on the amplitude.
 
\begin{figure}
\begin{center}
\epsfig{file=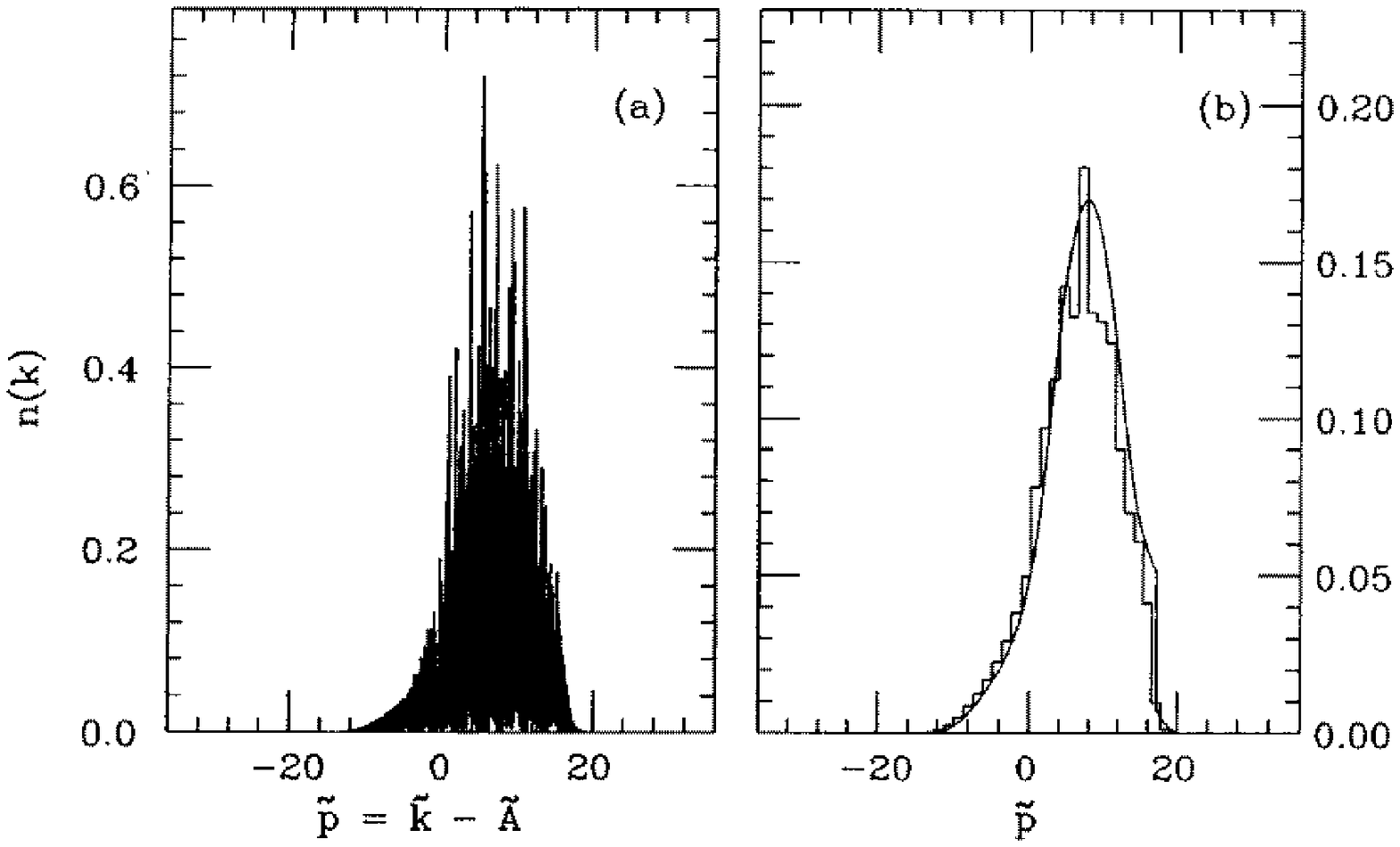,width=14cm}
\end{center}
\caption{(a) Momentum distribution of produced  pairs,  for  the  evolution
shown in Fig.~\ref{fig3_1}, at time $\tau=600.$  The abscissa  is  the  scaled  kinetic
momentum $\tilde p\equiv\tilde k-\tilde A$, with $\tilde k\equiv  k/m$.  (b)
Data of (a) after binning (histogram), compared with Boltzmann-Vlasov  model
(curve).
\label{fig3_3}}
\end{figure}

\begin{figure}
\begin{center}
\epsfig{file=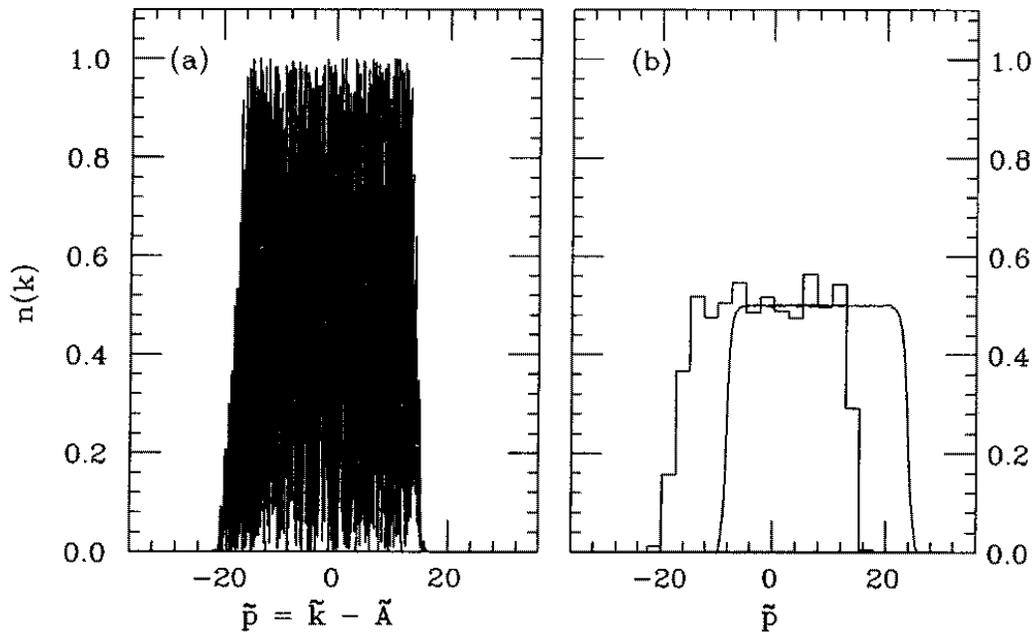,width=14cm}
\end{center}
\caption{The same as Fig.~\ref{fig3_3}, but for  the  evolution  shown  in  Fig.~\ref{fig3_2}
(i.e., initial field $\tilde E=4.0$), at time $\tau=200.$
\label{fig3_4}}
\end{figure}

We display in Figs.~\ref{fig3_3}(a) and~\ref{fig3_4}(a) the momentum distribution
of the produced particles, defined
by\footnote{Cf.~(\ref{bose_n}); $n$ here is an average over spin
states.  For the explicit form of $n(k;t)$ for large times,
analogous to (\ref{bose_density}), see \cite{PRD}.}
\begin{equation}
n(k;t)\equiv\frac{1}{2}\sum_{s=1,2}\left\langle
0\left\vert b^{(0)\dagger}_s({\bf  k};t)  b^{(0)}_s({\bf  k};t)
\right\vert 0\right\rangle\ .
\label{PRD_4.10}
\end{equation}
In accordance with the Pauli principle, the occupation number
does
not exceed one.
 
\subsubsection{Phenomenological transport equations}
 
The fermionic counterpart of (\ref{PRL_17}) is
\begin{eqnarray}
\frac{\partial f}{\partial t}+eE\frac{\partial f}{\partial p} &=&
\frac{dN}{dt\,dx\,dp}\nonumber\\
&=&-\bigl(1-2f(p,t)\bigr)\vert eE(t)\vert\nonumber\\
&&\quad\mbox{}\times\log
\left[1-\exp \left(-\frac {\pi m^2}{\vert
eE(t)\vert}\right)\right]
\, \delta (p).
\label{PRD_4.18}
\end{eqnarray}
Recall that $p$ is the {\it  kinetic}  momentum.  The
right-hand side is the Schwinger expression for the fermion
pair-production rate in  1+1  dimensions, multiplied by the
statistical factor $(1-2f)$ which represents Pauli blocking and
detailed balance (see the brief discussion above
Eq.~(\ref{PRL_17})).
 
Note that the singularity on the right-hand side of
(\ref{PRD_4.18})
for $ m=0 $ is spurious.
The Schwinger formula is based on a tunneling picture \cite{CNN}
that incorporates conservation of energy, but ignores other
conservation rules that would follow from specific features of
the system in question.  In the present case, where dynamical
photons are absent, chirality conservation actually {\em forbids}
pair production from the homogeneous electric field for massless
particles in one spatial dimension, so that the use of the
Schwinger formula is unjustified in this limit.
In the field theory, however, there is no such limitation.
The $m\to0$ limit is nothing other than the Schwinger model
\cite{Kogut,Schwinger61}, and the chiral anomaly gives a finite
rate \cite{Kogut,AGP83} for pair production.
 
Eq.~(\ref{PRD_4.18}) is solved in the same way as (\ref{PRL_17}).
The time evolutions of the scaled field strength $\tilde{E}$ and
current $\tilde\jmath$ are shown in the dashed curves of
Figs.~\ref{fig3_1}
and~\ref{fig3_2}.  The quantitative agreement between
the kinetic theory and the quantum theory is striking.
The amplitude of the electric field approaches a limiting value
after a few oscillations, meaning that thereafter the production
of particles is
negligible. In the boson case an analogous effect is seen,
but it sets in somewhat later than for
fermions.  The constant amplitude reflects the absence of pair
creation from virtual photons and the exponentially small spontaneous
pair creation rate at this stage of the evolution. The fact that the
electric field reaches its limiting value more quickly for fermions
than for bosons may be due to the difficulty of producing more
fermions once the low-momentum states have been occupied.
 
The comparison between $f(p,t)$ at a late time and
the smoothed $n(k)$  of  the
quantum theory is shown in Figs.~\ref{fig3_3}(b)  and~\ref{fig3_4}(b).\footnote{Again, the curves
have a relative displacement due to the slightly different  value
of $A$.}  The importance of Pauli blocking is obvious.
Without the statistical factor in (\ref{PRD_4.18})
the
occupation number exceeds one if the initial electric field is strong enough
\cite{PRD}.  The Schwinger source  term
in (\ref{PRD_4.18}) is very large for a strong electric field; it is  the
$(1-2f)$ term that prevents the violation of the Pauli principle.
 
As for the boson case, our comparison provides a justification for using
the kinetic theory in studying the quark-gluon plasma.
Past treatments have  not,  however,  included  the
Pauli-blocking term which is crucial for strong fields.
 
\section{Concluding remarks}
 
We have solved semiclassical QED for  strong  fields   using  the  adiabatic
regularization  method  for  spatially-homogeneous   electric   fields.   We
determine initial values of the  fields  and  their  derivatives  which  are
consistent both with the coupled Maxwell and matter-field equations and with
 the  adiabatic  regularization  scheme.    This   is    a   nonperturbative
calculation, which enables us to investigate  the dynamical evolution of the
interacting system of  matter and the  electromagnetic   field.   Careful
computational work was essential in order to  achieve   numerical  stability
for this system since it consists of three time scales  which are related to
frequencies of the order of the mass of the produced  particles (short  time
scale), the plasma oscillations of the electric  field and  current  (medium
time scale), and the degradation of the electric  field to small values,  by
which point particle  production  is  negligible   (long  time  scale).   It
requires fine grids in momentum  space and small time  steps  to  solve  the
differential equations.
 
Physical features like plasma oscillations and  the  plateau  in  the  first
period  of the current evolution  are  understood  by  solving  a  classical
relativistic  system  of particles in the presence  of  an  electric  field.
From the value of the plasma oscillation frequency  or  the  height  of  the
first plateau of  the current we obtain two separate estimates of the number
of particles at the  corresponding times.
 
There is agreement between QED and a simple  phenomenological model based on
classical Boltzmann-Vlasov equations  supplemented with a Schwin\-ger WKB
formula.   This agreement is significantly improved when the source term  is
multiplied   by   a   Pauli-blocking   or   Bose-enhancement   factor.   The
phenomenological model enables us to obtain immediate physical  insight  and
allows us  perform  calculations that are much faster and easier than  those
of the full QED formulation. In addition in the  phenomenological  model  we
understand the meaning of the number  of  particles,  whereas  in  QED  this
concept is ill-defined as long as the   interaction  takes  place.  The
agreement between the number of particles in these two  calculations  then
allows us to assess the quality of the  definition of the number  of
particles  in the field theory in the presence of interaction.
 
The existence of  a  phenomenological  model,  in this simplified situation,
with physical and numerical content very close to that of the exact
field-theory treatment opens  the  way  for  studies  that  include  spatial
dependence, real radiation, and, eventually, color degrees of  freedom.   At
the   same   time,   it   suggests   that   further   application   of
phenomenological transport equations for systems involving pair production
is in order.
\vskip 12pt
 
\noindent{\Large \bf Acknowledgements}\\
 
Major progress on this subject was made during visits of Y.K.  and  B.S.  to
the Theory Division at the Los Alamos National Laboratory.
We are very grateful  for having had the opportunity to work there  with
F.~Cooper and E.~Mottola. Their willingness to collaborate and to share their
physical insight aided this  work  substantially.  The  hospitality  of  the
Theoretical Division at Los Alamos National Laboratory during  these  visits
is also greatly appreciated, as is  the  access  the  Division  provided  to
computing facilities there.
 
We wish to thank I. Paziashvili for many enlightening discussions,  as  well
as A. Casher, M.S. Marinov, S. Nussinov, N. Weiss, and S. Yankielowicz for valuable
help.
 
Thanks are due to the German-Israel Foundation, the Minerva Foundation,  the
Alexander von Humboldt-Stiftung, and the Yuval Ne'eman
Chair in Theoretical Nuclear Physics at  Tel  Aviv  University  for  partial
support of this work.  The work of B.S. was partially supported by a
Wolfson Research Award administered by the Israel Academy of Sciences and
Humanities.
Part of this work was carried out while J.M.E. was  visiting
the Institut f\"ur Theoretische Physik der Universit\"at Frankfurt, to  whom
he is grateful for their warm hospitality.
 
\newcounter{mysec}
\newcommand{\myappendix}{\appendix
\setcounter{mysec}{0}
 \renewcommand{\themysec}{\Alph{mysec}}}
\newcommand{\myappsection}[1]{\section*{#1}
 \setcounter{equation}{0}
  \addtocounter{mysec}{1}}
\renewcommand{\theequation}{\themysec.\arabic{equation}}
 
\begin{myappendix}

\myappsection{Appendix: Particle spectra}

 
During the process of  particle  production,  the  particle  number  is  not
conserved, and we are not in an out-state region. However, it is possible to
introduce   an   interpolating   particle number   operator,    using    the
time-dependent  creation  and  annihilation  operators  of  the  first-order
adiabatic vacuum. This interpolating number operator has the  property  that
if at $t=0$ the initial state is equal to the  first-order adiabatic  vacuum
state, then  the  number  operator  starts  at  zero.  At  late  times,  the
first-order adiabatic number operator approaches the usual out-state  number
operator.
 
The wave functions of the first-order adiabatic expansion,
\begin{eqnarray}
\phi^{(0)}_{\bf{k}}({\bf{x}},t) & = & e^{i {{\bf{k}} \cdot {\bf{x}}}}
 f^{(0)}_{\bf{k}}(t), \nonumber\\
f^{(0)}_{\bf{k}} (t) &=& {{e^{-i \int_0^t \omega_{\bf{k}}(t^{\prime})
 dt^{\prime}}} \over
{(2 \omega_{\bf{k}}(t))^{1/2}} }
= {e^{-iy^{(0)}_{\bf{k}}(t)} \over {(2 \omega_{\bf{k}}(t))^{1/2}}}, \nonumber
\\
\bar{\phi}^{\ast(0)}_{\bf{k}}({\bf{x}},t)
 &=&  f_{-{\bf{k}}}^{\ast (0)}(t) e^{-i{\bf{k}} \cdot {\bf{x}}} ,
 \label{s_A1}
\end{eqnarray}
form an alternative basis for expanding the quantum field $\Phi$, so that we
can write
\begin{equation}
\Phi(t,z,{\bf{x}}_{\perp}) = \int [d{\bf{k}}]
 \left(a_{\bf{k}}(t) \phi^{(0)}_{\bf{k}}({\bf{x}},t)
+ b^{\dagger}_{\bf{k}}(t)\bar{\phi}^{\ast (0)}_{\bf{k}}({\bf{x}},t) \right).
\label{s_A2}
\end{equation}
Previously we expressed the field in terms of time-independent creation and
annihilation operators:
\begin{equation}
\Phi(t,z,{\bf{x}}_{\perp}) = \int [d{\bf{k}}]
 \left( a_{{k}}\phi_{\bf{k}}({\bf{x}},t)
+ b^{\dagger}_{\bf{k}}\bar{\phi}^{\ast}_{\bf{k}}({\bf{x}},t) \right)
 \label{s_A3}
\end{equation}
and
\begin{equation}
\phi_{\bf{k}}({\bf{x}},t) = f_{\bf{k}}(t) e^{i {{\bf{k}} \cdot {\bf{x}}}},
\quad
\bar{\phi}^{\ast}_{\bf{k}}({\bf{x}},t) = f_{-{\bf{k}}}^{\ast}(t)
e^{-i {{\bf{k}}\cdot {\bf{x}}}}.
\label{s_A4}
\end{equation}
Because both $f$ and $f^{(0)}$ obey the Wronskian condition, it follows that
\begin{eqnarray}
\left(\phi_{\bf{k}}({\bf{x}},t),\phi_{{\bf{k}}^{\prime}}({\bf{x}},t) \right)
 &=&
\left(\phi^{(0)}_{\bf{k}}({\bf{x}},t),\phi^{(0)}_{{\bf{k}}^{\prime}}({\bf{x}},
t)
 \right)
 = (2\pi)^3 \delta^3 \left( {\bf{k}} -{\bf{k}}^{\prime}\right), \nonumber \\
\left(\phi^{\ast}_{\bf{k}}({\bf{x}},t),\phi_{{\bf{k}}^{\prime}}({\bf{x}},t)
 \right) &=&
\left(\phi^{\ast
 (0)}_{\bf{k}}({\bf{x}},t),\phi^{(0)}_{{\bf{k}}^{\prime}}({\bf{x}},t) \right)
=0,
\label{s_A5}
\end{eqnarray}
where the inner product is defined as
\begin{equation}
(u,v) \equiv i \int d^{3} x \left( u^{\ast} \frac{\partial v} {\partial t}
- \frac{\partial u^{\ast}} {\partial t} v \right).
\label{s_A6}
\end{equation}
Using this we find that
\begin{equation}
a_{\bf{k}} = \left(\phi_{\bf{k}}({\bf{x}},t),\phi ({\bf{x}},t) \right),
 \quad
b^{\dagger}_{\bf{k}} = - \left( {\bar{\phi}}^{\ast}_{\bf{k}}({\bf{x}},t).
\phi ({\bf{x}},t).
\right)
\label{s_A7}
\end{equation}
 
At large times $t$, the time variation in the  time-dependent  creation  and
annihilation  operators  connected  with  the  first-order  adiabatic   wave
functions  becomes small. In that regime (which is not quite the out regime)
we can determine these operators through the relations
\begin{equation}
a_{\bf{k}}(t) = \left(\phi^{(0)}_{\bf{k}}({\bf{x}},t),\phi ({\bf{x}},t)
 \right),
\quad
b^{\dagger}_{\bf{k}}(t) = - \left({\bar{\phi}}^{\ast
 (0)}_{\bf{k}}({\bf{x}},t),\phi
({\bf{x}},t) \right).
\label{s_A8}
\end{equation}
From these we can explicitly evaluate the Bogolyubov transformation at large
times. In general we have
\begin{equation}
a_{\bf{k}} (t) = \alpha({\bf{k}},t) a_{\bf{k}} + \beta^{\ast}({\bf{k}},t)
 b^{\dagger}_{\bf{k}},
\quad
b_{\bf{k}}(t) = \alpha({\bf{k}},t) b_{\bf{k}} + \beta^{\ast}({\bf{k}},t)
 a^{\dagger}_{\bf{k}}.
\label{s_A9}
\end{equation}
At late times, these operators become time-independent, and $\alpha$
and $\beta $ are given by
\begin{eqnarray}
\alpha({\bf{k}},t)& = &i \left( f^{(0)\ast}_{\bf{k}} \frac {\partial
 f_{\bf{k}}}{\partial t} - \frac {\partial f^{(0)\ast}_{\bf{k}}}{\partial t}
 f_{\bf{k}} \right),
\nonumber \\
\beta({\bf{k}},t))& = &i \left( f^{(0)}_{\bf{k}} \frac {\partial
 f_{\bf{k}}}{\partial t}
- \frac {\partial f^{(0)}_{\bf{k}}}{\partial t} f_{\bf{k}} \right).
\label{s_A10}
\end{eqnarray}
 
The time-dependent interpolating particle number operator is defined by
\begin{equation}
N({\bf{k}},t) (2 \pi)^3 \delta^3 ({\bf{k}} -{\bf{k}}^{\prime}) =
\langle a^{\dagger}_{\bf{k}}(t) a_{\bf{k}}(t) \rangle,
\label{s_A11}
\end{equation}
where $N_{+} =N_{-} = N.$ Thus,
\begin{equation}
N({\bf{k}},t) = N({\bf{k}}) | \alpha({\bf{k}},t) |^{ 2} +(1+N({\bf{k}}))
 |\beta({\bf{k}},t)|^{2}
 2 Re\{\alpha({\bf{k}},t)\beta({\bf{k}},t) F({\bf{k}})\} .
\label{s_A12}
\end{equation}
For the case of  $N=F=0$ we obtain at late times
\begin{equation}
N({\bf{k}},t) = {\frac {1} {4 \omega_{\bf{k}}\Omega_{\bf{k}}}}
\left[ (\Omega_{\bf{k}}-\omega_{\bf{k}})^2 + {\frac {1}{4}}
\left( \frac {\dot{\Omega}_{\bf{k}}} {\Omega_{\bf{k}}} -
\frac {\dot{\omega}_{\bf{k}}} {\omega_{\bf{k}}} \right)^{2} \right].
\label{s_A13}
\end{equation}
We note that our initial conditions ensure that initially this
interpolating number operator is zero.
\end{myappendix}


\begin{thebibliography}{99}

 
\bibitem{Parker68}
L. Parker, Phys. Rev. Lett. {\bf 21}, 562 (1968).
 
\bibitem{Parker69}
L. Parker, Phys. Rev. {\bf 183}, 1057 (1969).
 
\bibitem{Parker71}
L. Parker, Phys. Rev. D {\bf 3}, 346 (1971).
 
\bibitem{Zel71}
Ya.~B. Zel'dovich and A. A. Starobinsky, Zh. Eksp. Teor.  Fiz.  {\bf  26},
2161 (1971) [Sov. Phys. JETP {\bf 34}, 1159 (1972)].
 
\bibitem{Parker74}
L. Parker and S. A. Fulling, Phys. Rev. D {\bf 9}, 341 (1974).
 
\bibitem{Fulling74}
S. A. Fulling, L. Parker, and B. L.  Hu,  Phys.  Rev.  D  {\bf  10},  3905
(1974); Phys. Rev. D {\bf 11}, 1714 (1975).
 
\bibitem{Bunch78}
T. S. Bunch, S. M. Christensen, and S. A. Fulling, Phys. Rev. D  {\bf  18},
4435 (1978).
 
\bibitem{Hu78}
B. L. Hu, Phys. Rev. D {\bf 18}, 4460 (1978).
 
\bibitem{Birrell78}
N. D. Birrell, Proc. Roy. Soc. A {\bf 361}, 513 (1978).
 
\bibitem{Hu79}
B. L. Hu, Phys. Lett. A {\bf 71}, 169 (1979).
 
\bibitem{Bunch80}
T. S. Bunch, J. Phys. A {\bf 13}, 1297 (1980).
 
\bibitem{Suen87a}
W.-M. Suen, Phys. Rev. D {\bf 35}, 1793 (1987).
 
\bibitem{Suen87b}
W.-M. Suen and P. R. Anderson, Phys. Rev. D {\bf 35}, 2940 (1987).
 
\bibitem{Lieberman88}
B. Lieberman and B. Rogers, Phys. Rev. D {\bf 38}, 3648 (1988).
 
\bibitem{Suen89}
W.-M. Suen, Phys. Rev. D {\bf 40}, 315 (1989).
 
\bibitem{CM}
F. Cooper and E. Mottola, Phys. Rev. D {\bf 40}, 456 (1989).
 
\bibitem{Rogers90}
B. Rogers, Phys. Rev. D {\bf 42}, 2069 (1990).
 
\bibitem{BD}
N. D. Birrell and P. C. W. Davies, {\it Quantum  Fields  in  Curved
Space} (Cambridge University Press, Cambridge, 1982).
 
\bibitem{Fulling}
S. A. Fulling, {\it Aspects of Quantum Field Theory in  Curved
Space-Time} (Cambridge University Press, Cambridge, 1989).
 
\bibitem{Zel72}
Ya.~B. Zel'dovich, in {\it Magic Without Magic:  John Archibald Wheeler,}
edited by J.~Klauder (Freeman, San Francisco, 1972), p.~277.
 
\bibitem{DeWitt}
B. S. DeWitt, Phys. Rep. {\bf 19}, 295 (1975).
 
\bibitem{Sauter}
F. Sauter, Z. Phys. {\bf 69}, 742 (1931).
 
\bibitem{HE}
W. Heisenberg and H. Euler, Z. Phys. {\bf 98}, 714 (1936).
 
\bibitem{Schwinger51}
J. Schwinger, Phys. Rev. {\bf 82}, 664 (1951).
 
\bibitem{Frogs}
E. Brezin and C. Itzykson, Phys. Rev. D {\bf 2}, 1191 (1970).
 
\bibitem{IZ}
C. Itzykson and J.-B. Zuber, {\it Quantum Field Theory}  (McGraw-Hill,
New York, 1980).
 
\bibitem{GMR}
W. Greiner, B. M\"uller,  and  J. Rafelski, {\it  Quantum
Electrodynamics of Strong Fields} (Springer, Berlin, 1985).
 
\bibitem{Kogut}
A. Casher, J. Kogut, and L. Susskind, Phys. Rev. D  {\bf  10},  732
(1974).
 
\bibitem{LowNussinov}
F. E. Low, Phys. Rev. D {\bf 12}, 163 (1975);
 
S. Nussinov, Phys. Rev. Lett. {\bf 34}, 1286 (1975).
 
\bibitem{CNN}
A. Casher, H. Neuberger, and S. Nussinov, Phys. Rev. D {\bf 20}, 179
(1979);
 
H. Neuberger, {\em ibid.} {\bf 20}, 2936 (1979);
 
A. Casher, H. Neuberger, and S. Nussinov, {\em ibid.} {\bf 21}, 1966
(1980).
 
\bibitem{Lund}
B. Andersson, G. Gustafson, G. Ingelman, and T. Sjostrand, Phys. Rep.
{\bf 97}, 31 (1983).
 
\bibitem{Glendenning83}
N. K. Glendenning and T. Matsui, Phys. Rev. D {\bf 28}, 2890 (1983);
 
B. Banerjee, N. K. Glendenning, and T. Matsui, Phys. Lett. {\bf 127B},
453 (1983).
 
\bibitem{Marinov77}
M. S. Marinov and V. S. Popov, Fortsch. Phys. {\bf25}, 373 (1977).
 
\bibitem{Grib}
A. A. Grib, S. G. Mamaev and V. M. Mostepanenko, {\it Quantum  Effects
in Strong External Fields} (Atomizdat, Moscow, 1980) (in Russian);
 
N. B. Narozhnyi and A. I. Nikishov, Yad. Fiz. {\bf 11},  1072  (1970)
[Sov. J. Nucl. Phys. {\bf 11}, 596 (1970)];
 
V. M. Mostepanenko, Yad. Fiz. {\bf 30},  208  (1979)
[Sov. J. Nucl. Phys. {\bf 30}, 107 (1979)];
 
S. G. Mamaev and N. N. Trunov, Yad. Fiz. {\bf 30},  1301  (1979)
[Sov. J. Nucl. Phys. {\bf 30}, 677 (1979)].
 
\bibitem{finvol}
R.-C. Wang and C.-Y. Wong, Phys. Rev. D {\bf 38}, 348 (1988);
 
C. Martin and D. Vautherin, Phys. Rev. D {\bf 38}, 3593 (1988);
 
C. Martin and D. Vautherin, Phys. Rev. D {\bf 40}, 1667 (1989);
 
Th. Sch\"onfeld, A. Sch\"afer, B. M\"uller, K. Sailer, J. Reinhardt,
and W. Greiner, Phys. Lett. B {\bf 247}, 5 (1990);
 
H.-P. Pavel and D.M. Brink, Z. Phys. C {\bf 51}, 119 (1991);
 
C.S. Warke and R.S. Bhalerao, Pramana J. Phys. {\bf 38}, 37 (1992).
 
\bibitem{Ambjorn83}
J. Ambj\o rn and S. Wolfram, Ann. Phys. (NY){\bf 147}, 33 (1983).
 
\bibitem{Muller}
B. M\"uller, {\it The Physics  of  the  Quark-Gluon  Plasma}  (Springer,
Berlin, 1985).
 
\bibitem{Hwa}
R. Hwa, ed., {\it Quark-Gluon Plasma} (World Scientific, Singapore,
1990).
 
\bibitem{BC86}
A. Bia\l as and W. Czy\.z, Phys. Rev. D {\bf 30}, 2371 (1984);
{\em ibid.} {\bf 31}, 198 (1985);
Z. Phys. {\bf C28}, 255 (1985);
Nucl. Phys. {\bf B267}, 242 (1985);
Acta Phys. Pol. B {\bf17}, 635 (1986).
 
\bibitem{BC88}
A. Bia\l as, W. Czy\.z, A. Dyrek, and W. Florkowski, Nucl. Phys. {\bf B296},
611 (1988).
 
\bibitem{Kajantie85}
K. Kajantie and T. Matsui, Phys. Lett {\bf 164B}, 373 (1985).
 
\bibitem{Gatoff87}
G. Gatoff, A. K. Kerman, and T. Matsui, Phys. Rev. D {\bf 36}, 114 (1987).
 
\bibitem{Dutch}
S. R. De Groot, W. A. van Leeuwen, and Ch.~G. van Weert, {\it
Relativistic Kinetic Theory} (North-Holland, Amsterdam, 1980).
 
\bibitem{Heinz83}
U. Heinz, Phys. Rev. Lett. {\bf 51}, 351 (1983).
 
\bibitem{Gyu85}
M. Gyulassy and A. Iwazaki, Phys. Lett. B {\bf 164}, 157 (1985).
 
\bibitem{Elze86}
H.-Th. Elze, M. Gyulassy and D. Vasak, Nucl. Phys. {\bf B276}, 706 (1986).
 
\bibitem{JME88}
J. M. Eisenberg and G. K\"albermann, Phys. Rev. D {\bf 37}, 1197 (1988).
 
\bibitem{Poles}
I. Bialynicki-Birula, P. G\'ornicki, and J. Rafelski, Phys. Rev. D {\bf
44}, 1825 (1991).
 
\bibitem{prep}
C. Best and J. M. Eisenberg, University of  Frankfurt  preprint,  September,
1992;
 
S. Graf, University of  Frankfurt  preprint,  September, 1992;
 
Y. Kluger, in preparation.
 
\bibitem{PRL}
F. Cooper, E. Mottola, B. Rogers, and P. Anderson, in
{\it Intermittency in High Energy Collisions,} edited by F. Cooper,
R. C. Hwa, and I. Sarcevic (World Scientific, Singapore, 1991), p. 399;
 
Y. Kluger, J. M. Eisenberg, B. Svetitsky, F. Cooper, and E. Mottola,
Phys. Rev. Lett. {\bf 67}, 2427 (1991).
 
\bibitem{PRD}
Y. Kluger, J. M. Eisenberg, B. Svetitsky, F. Cooper, and E. Mottola,
Phys. Rev. D {\bf 45}, 4659 (1992).
 
\bibitem{CooperFrye}
F. Cooper, G. Frye, and E. Schonberg, Phys. Rev. D {\bf 11}, 192 (1975).
 
\bibitem{Bj83}
J.D. Bjorken, Phys. Rev. D {\bf 27}, 140 (1983).
 
\bibitem{biv}
F. Cooper, J. M. Eisenberg, Y. Kluger, E. Mottola, and  B.
Svetitsky, Tel Aviv preprint TAUP 1944-92, November 1992 (unpublished);
 
F. Cooper, Lectures given at the NATO ASI on Particle Production in
Highly-Excited Matter, Il Ciocco, Italy, July, 1992, Los Alamos preprint
LA-UR-92-2753, August, 1992 (unpublished).
 
\bibitem{Biro84}
T. S. Bir\'{o}, H. B. Nielsen, and J. Knoll, Nucl. Phys. {\bf B245}, 449
(1984).
 
\bibitem{Kerman86}
A. K. Kerman, T. Matsui, and B. Svetitsky, Phys. Rev. Lett. {\bf 56},
219 (1986).
 
\bibitem{Waterman73}
P. C. Waterman, Am. J. Phys. {\bf 41}, 373 (1973).
 
\bibitem{Trafton71}
L. Trafton, J. Comp. Phys. {\bf 8}, 64 (1971).
 
\bibitem{BjD}
J. D. Bjorken and S. D. Drell, {\it Relativistic Quantum Mechanics,}
(McGraw-Hill, New York, 1964).
 
\bibitem{GMF}
A. A. Grib, V. M. Mostepanenko, and V. M. Frolov, Teor. Mat.
Fiz. {\bf 13}, 377 (1972) [Theor. Math. Phys. {\bf 13}, 1207 (1972)].
 
\bibitem{Schwinger61}
J. Schwinger, Phys. Rev. {\bf 128}, 2425 (1961).
 
\bibitem{AGP83}
J.~Ambj\o rn, J.~Greensite, and C.~Peterson, Nucl. Phys. {\bf B221}, 381
(1983).
 
\end{thebibliography}
\end{document}